\documentclass[amsmath,amssymb,aps,prd,11pt,tightenlines,superscriptaddress, 
nofootinbib,preprintnumbers,notitlepage]{revtex4-1}
\usepackage{graphicx}  
\usepackage{dcolumn}   
\usepackage{bm}        
\usepackage{amssymb}   
\usepackage{amsmath}
\usepackage{mathrsfs}  
\usepackage{graphicx,color} 
    
\usepackage[english]{babel}     
\usepackage{graphicx,wrapfig,lipsum}      
\usepackage{multimedia}    
\usepackage[mathscr]{eucal}  
\usepackage{bm}               
\usepackage{amssymb}        
\usepackage{amsmath}     
\usepackage{mathrsfs}  
\usepackage[mathscr]{eucal}

\begin{document}     

\title{4d ensembles of percolating center vortices and monopole defects: \\ the emergence of flux tubes with $N$-ality  and gluon confinement }   
 
    
\author{L.~E.~Oxman }  
  
\affiliation{
Instituto de F\'\i sica, Universidade Federal Fluminense,   
24210-346 Niter\'oi - RJ, Brasil.} 

\date{\today}

\begin{abstract}     
 
Ensembles of magnetic defects represent {\it quantum} variables that have been detected and extensively explored in lattice ${\rm SU}(N)$ pure Yang-Mills theory. They successfully explain many properties of confinement and are strongly believed to capture the (infrared)  path-integral measure. In this work, we initially motivate the presence of magnetic non-Abelian degrees of freedom in these ensembles.  Next, we consider a simple Gaussian model to account for fluctuations. In this case, both center vortices and monopoles become relevant degrees 
in Wilson loop averages.  These physical inputs are then implemented in an ensemble of percolating center vortices in four dimensions by proposing a measure 
to compute center-element averages. The introduction of phenomenological information such as monopole tension, stiffness, and fusion leads to an effective YMH model with adjoint Higgs fields. 
If monopoles also condense, then the gauge group undergoes ${\rm SU}(N) \to {\rm Z}(N)$ SSB. This pattern has been proposed as a strong candidate to describe confinement. In the presence of external quarks, these models are known to be dominated by {\it classical} solutions, formed by flux tubes with $N$-ality as well as by confined dual monopoles (gluons).

\end{abstract}  
 

\maketitle

\section{Introduction} 
       
Lattice simulations have consistently established that a confining linear potential between a fundamental quark and antiquark is generated in the infrared regime of pure ${\rm SU}(N)$ Yang-Mills theory \cite{Bali}. For other quark representations, the asymptotic potential depends solely on how the center of ${\rm SU}(N)$ is realized  \cite{dFK}.  This is one of the properties that favors a quark confinement mechanism based on an ensemble of center vortices  \cite{tHooft:1977nqb}-\cite{Nielsen:1979xu}. When the quark Wilson loop is linked by a center vortex, it gains a center element. Thus,   in the percolating phase, the area law obtained naturally displays $N$-ality.  
This idea has gained momentum over the last many years, settling these degrees as essential infrared quantum variables to  capture the path-integral measure \cite{Deb+97}-\cite{Gattnar:2004gx}.   
On the other hand,  Monte Carlo simulations also show subleading contributions that coincide with universal L\"uscher corrections due to the transverse (quantum)  fluctuations of a string \cite{LW}.  Moreover, the action and field distributions measured around the confining string are nontrivial, revealing  a chromoelectric flux tube \cite{su2-Bali-1995}-\cite{Cosmai-2017}.  
While center vortices are essential to describe an area law with $N$-ality,  L\"uscher terms and flux tubes have not yet been observed in these ensembles.
 
In contrast, dual superconductivity \cite{tHooft:1981bkw}-\cite{Baker-Dosch} is suitable to accomodate stringlike behavior. The idea of Abelian projection \cite{tHooft:1981bkw} and associated ensembles of monopole defects were analyzed in the lattice \cite{mon-ref1}-\cite{mon-ref3}.  The understanding of confinement in compact QED, as well as the manner through which the proliferation of monopoles induce observable surfaces attached to a quark loop, was obtained in Refs. \cite{polya}, \cite{Polya97}.  In addition, the profile of the confining Yang-Mills flux tube  has been fitted using vortex solutions in effective Abelian Higgs models \cite{su2-Bali-1995}-\cite{Cosmai-2017}. However, Abelian scenarios cannot describe $N$-ality. For example,  when applied to double Wilson loops in ${\rm SU}(2)$, they lead to the sum of areas, instead of the difference-in-areas law observed in the lattice  and accomodated by center vortices \cite{GR}.       
  
Based on the complementary properties of center vortices and monopole defects, it is natural to infere that an appropriate combination of both could capture the whole physical picture.  
Indeed, in lattice calculations of pure ${\rm SU}(N)$ Yang-Mills (YM) theory based on center
 gauges,  both center vortices and attached monopoles were detected, forming chains. In fact, they account for $97 \%$ of the cases  \cite{Ambjorn:1999ym}. In the continuum, the description and topological aspects of these arrays, in which the Lie algebra flux orientation changes at the monopoles, were worked out in Ref. \cite{Reinhardt:2001kf}. In Abelian gauges, the possibility that integrating off-diagonal fluctuations could lead to collimated chains was suggested in Ref. \cite{Greensite-conf}, and references therein.

Another scenario to accommodate $N$-ality has been proposed at the level of possible dual descriptions.  The properties of ${\rm SU}(N)$ YM confining strings have been sought in classical topological solutions by exploring a variety of models. Flux tubes with $N$-ality and  confined dual monopoles are known to exist in ${\rm SU}(N)$ Yang-Mills-Higgs (YMH) models when the gauge group is spontaneously broken to ${\rm  Z}(N)$ \cite{Fidel2}-\cite{Auzzi-Kumar}. Another possibility to  accommodate these states is provided by non-supersymmetric models with $N$ fundamental Higgs fields \cite{GSY}, ${\rm SU}(N)\times {\rm U}(1)$ gauge group, and a color-flavor locking phase that equips flux tubes and dual monopoles with a non-Abelian moduli space \cite{David}-\cite{David-12}.
 The connection of color-flavor locking with monopole condensates that fit into the Goddard-Nyuts-Olive classification scheme \cite{GNO}   
was extensively analyzed, mainly in a supersymmetric context. The present status can be found in Ref. \cite{Konishi-2018}.  A  color-flavor locking phase could also be present in ${\rm SU}(N)$ YMH models with $N^2-1$ (real) adjoint Higgs fields \cite{conf-qg}-\cite{LD}.  Confined dual monopoles were interpreted as gluons in Refs. \cite{GSY}, \cite{Shifman-Yung-gluon},  \cite{Marco-glue} (see also \cite{conf-qg}, \cite{Proceed}). 
   
The aim of this work is to  combine the different ideas  into a possible unified mechanism. 
Chains were visualized as magnetic defects of a local color frame in Ref. \cite{lucho}. Their relation to the observability of surfaces attached to the quark loop was discussed in Ref. \cite{l-dual}. In Ref. \cite{deLemos:2011ww}, the derivation of a 3d effective field model for chains  
made it possible to relate the monopole (instanton) component with the ${\rm  Z}(N)$-symmetric terms in the 't Hooft  model \cite{tHooft:1977nqb}.  Therefore, when center vortices condense, monopoles are essential to drive magnetic ${\rm Z}(N)$ SSB and generate an observable domain wall with $N$-ality, attached to the quark loop. A generalized non-Abelian model to describe phases with different vortex pairings was introduced in Ref. \cite{F-O}. The relation between 4d ensembles of monopoles that carry adjoint charges and models based on a set of adjoint Higgs fields was suggested in Ref. \cite{conf-qg}. This idea was further elaborated in  Ref. \cite{GBO}, where we applied polymer techniques to an ensemble of worldlines  with non-Abelian d.o.f.   

In four dimensions, while monopoles are naturally described by effective field models \cite{bar-sam}-\cite{HSi}, the consideration of center vortices would be related to {\it string} field theories, which poses important difficulties.  We suggest that when center vortices percolate,  the effect of linking numbers could be captured in an effective {\it field} theory. This is motivated  by the low-energy effective description of higher dimensional defect condensates \cite{Rey}-\cite{Gri2010}, the results of a recent study about ensembles of center vortices in 3d \cite{Oxman-Reinhardt-2017}, and a simple model based on a smoothed Gaussian Wilson loop.
 
In sections \ref{detecting} and \ref{nonAdof}, based on a gauge fixing in the continuum \cite{OS-det} that is  motivated by lattice center gauges \cite{VW}-\cite{faber-2}, we show the presence of non-Abelian d.o.f. in configurations with 
 center vortices and monopole defects.   In section \ref{simpli}, relying on the Petrov-Diakonov representation, we present a simple example in which center vortices and  monopoles with non-Abelian d.o.f. have a combined effect on Wilson loop averages. An ensemble measure that mixes percolating center vortices and chains is then proposed in section \ref{first}. In sections \ref{frules} and \ref{eff-fields}, fusion rules between monopole  adjoint lines are associated with effective Feynman diagrams, and the  ensemble partition function is rewritten in terms of a dual ${\rm SU}(N)$ YMH model. Finally, in section \ref{conc}, we present our conclusions.    
   
\vspace{.2cm} 

Throughout this work, we shall use the internal product between a pair of Lie algebra elements $X, Y \in \mathfrak{su}(N)$
\begin{equation} 
( X, Y ) = {\rm tr}\, ({\rm Ad}(X) {\rm Ad}(Y) )  \;,
\label{Kf} 
\end{equation}
where ${\rm Ad }(\cdot) $ refers to the adjoint representation, and shall denote $(X, X) \equiv (X)^2 $. The main properties of this product are the cyclic and group invariances, which are a consequence of the defining property of a representation,
\begin{gather}
(X, [Y,Z]) = (Z, [X,Y])  \makebox[.5in]{,} ( UXU^{-1}, UYU^{-1} ) = ( X, Y )  \;,  \\ 
{\rm Ad}([X,Y]) = [{\rm Ad}(X) , {\rm Ad}(Y) ] \makebox[.3in]{,} {\rm Ad}(UXU^{-1}) = R(U) {\rm Ad}(X) R^{-1}(U) \;.
\end{gather}
 $R(U)= {\rm Ad}(U)$ is the $\mathscr{D}_{\rm Ad} \times \mathscr{D}_{\rm Ad}$ matrix that represents $U$ in the adjoint ($\mathscr{D}_{\rm Ad} = N^2-1$). We shall also adopt an orthonormal Lie basis $T_A$, $A=1, \dots, N^2-1$,
\begin{gather} 
( T_A,T_B ) = \delta_{AB}  \makebox[.5in]{,}  [T_A,T_B]=if_{ABC} T_{C} \;,
\label{Lie-alg}  \\ 
  {\rm Ad}(T_A)|_{BC}=-i f_{ABC} \makebox[.5in]{,}  f_{ABC}\, f_{DBC} =\delta_{AD} \;.
\label{conv}   
\end{gather}  
Matrices such as $U$, with no explicit reference to the irrep,  are understood to be in the  fundamental representation of ${\rm SU}(N)$.

\section{Detecting magnetic defects in the continuum}  
\label{detecting} 
   
In the lattice, gauge fixings designed to avoid the Gribov problem and detect center vortices were proposed in refs. \cite{VW}-\cite{faber-2} (for a review, see ref. \cite{greensite-livro}). They are based on the lowest eigenfunctions $(f_1, f_2, \dots)$  of the adjoint covariant Laplacian 
\begin{equation} 
D_{\mu}D_{\mu}(\mathscr{A})\, f_I = \lambda_I \, f_I \makebox[.5in]{,} D_{\mu}(\mathscr{A}) = \partial_{\mu} -i\,[\mathscr{A}_{\mu}, \;]   \;,
\label{covader}
\end{equation} 
using them to fix a prescribed orientation in color space. For example, in the direct Laplacian center gauge \cite{faber,faber-2},  a map ${\rm Ad}(S)$ is  constructed in a covariant way, that is, under a chromoelectric gauge transformation $\mathscr{A}_\mu^{U_{\rm e}}$, the associated map is ${\rm Ad} (U_{\rm e} S)$.  
For $N=2$, this is obtained from the polar decomposition of the real $3\times 3$ matrix formed by the color entries of $(f_1, f_2, f_3)$. 
Since this procedure is based on the lowest eigenfunctions, it cannot be directly implemented in the continuum. However, in Ref. \cite{OS-det}, we introduced a modified version where the assignment
\begin{equation}
\mathscr{A}_\mu \to f_I  \to {\rm Ad}(S) 
\label{mapping}
\end{equation}
is based on the adjoint fields $f_I \in \mathfrak{su}(N)$ that solve a set of coupled differential equations
\begin{equation}
\frac{\delta 
S_{\rm aux }}{\delta f_{I}}  = D_{\mu}D_{\mu}(\mathscr{A})\, f_I +  \dots 
= 0   \;.
\label{eqom}  
\end{equation}
 In order for $(f_1, f_2, \dots)$ to be strongly correlated with ${\rm Ad}(S)$, the auxiliary action $S_{\rm aux}$ possesses ${\rm SU}(N) \to {\rm Z}(N)$ SSB. 
 Considering $N^2-1$ fields, $I =1, \dots, N^2-1$,  the desired map was extracted from a polar decomposition of the tuple $(f_1, f_2, \dots )$ in terms of ``modulus'' $(q_1, q_2, \dots)$ and ``phase'' $S$-variables,
\begin{equation}
f_I = Sq_IS^{-1} \makebox[.5in]{,}   \sum_I [q_I , T_I] = 0 \;. 
\label{pmod}
\end{equation} 
The last condition amounts to looking for the rotated $f_I$'s that form the tuple which minimizes
\[ 
\sum _I (q_I -v T_I )^2 \;,
\]
where $(v T_1 , v T_2, \dots)$ is a prescribed point in the vacuum manifold of $S_{\rm aux}$. 
For ${\rm SU}(2)$, this makes contact with the polar decomposition of a real $3\times 3$ matrix. Again, because of covariance, for the gauge transformed field
\[
\mathscr{A}_\mu^{U_{\rm e}}  = U_{\rm e} \mathscr{A}_\mu \,U_{\rm e}^{-1} +   i\,  U_{\rm e}\partial_\mu U_{\rm e}^{-1}  \;,
\]
the extrated phase is $U_{\rm e}S $,
\begin{equation} 
\mathscr{A}_\mu^{U_{\rm e}} \to U_{\rm e} f_I \, U_{\rm e}^{-1}  \to {\rm Ad}(U_{\rm e} S) = {\rm Ad}(U_{\rm e}) \, {\rm Ad}(S) \;.
\label{t-map}
\end{equation} 
Although $\mathscr{A}_\mu$ is a well-defined variable, $S$ could contain defects. Therefore, the equivalence relation given by
\begin{equation} 
S \sim S' \makebox[.4in]{\rm if}  S'= U_{\rm e}S \;, ~~ {\rm regular}~ U_{\rm e}   \;, 
\label{i-rel}
\end{equation} 
induces a nontrivial partition of mappings into classes $[S]$, and of configurations into 
sectors $\mathscr{V}(S)$: two variables $\mathscr{A}_\mu$, $ \mathscr{A}'_\mu $ are in the same sector if they are mapped to $S$, $S'$ that are equivalent in the sense given by Eq. \eqref{i-rel}. 
This should not be confused with the equivalence relation
\begin{equation}   
\mathscr{A}_\mu \sim \mathscr{A}'_\mu \makebox[.4in]{\rm if}  \mathscr{A}'_\mu = \mathscr{A}_\mu^{U_{\rm e}} \;, ~~ {\rm regular}~ U_{\rm e}   \;.
\end{equation} 
In each sector $\mathscr{V}(S)$, there are infinitely many physically inequivalent configurations. For example, there is a perturbative sector formed by those $\mathscr{P}_\mu$ mapped to a regular $S$. Other sectors will be related to mappings with different numbers, types, and locations of defects. Equivalence classes of mappings will be denoted by $[S_0]$, where the label $S_0$ 
refers to a choice of representative. The gauge-fixed variables in 
$[S_0]$ satisfy
\[
A_\mu \to   {\rm Ad}(S_0) \;.
\] 
In the perturbative sector, $S_0$ can be chosen as the identity map, and the gauge-fixed perturbative variables satisfy, $P_\mu \to  I $. 
The total partition function for the YM theory
\begin{eqnarray}  
S_{\rm YM} = \int d^4x\, \frac{1}{4g^2_{\rm e}} \, (F_{\mu \nu} (\mathscr{A}))^2 \makebox[.5in]{,} 
F_{\mu \nu} (\mathscr{A})= \partial_\mu \mathscr{A}_\nu - \partial_\nu \mathscr{A}_\mu -i [\mathscr{A}_\mu , \mathscr{A}_\nu] \;,
\label{nona-f}
\end{eqnarray}  
is a sum over sectors $Z_{YM} = \sum_{S_0} \, Z_{YM}^{(S_0)} $, 
where $Z_{YM}^{(S_0)}$ are the gauge-fixed partial contributions. They are obtained from the path integral over $\mathscr{V}(S)$, $S = U_{\rm e} S_0$, by using an identity to introduce the equations of motion \eqref{eqom},  
\begin{equation}   
1 =\int [\mathcal{D} f_I]\; \delta\left(\frac{\delta  
S_{\rm aux }}{\delta f_{I}}\right)   \det \left( \frac{\delta^{2}S_{\rm aux}}{\delta 
f_{I}\, \delta f_{J}}\right) \;,
\label{ident}
\end{equation}   
then changing to polar variables $q_I$, and finally factorizing the regular part $U_{\rm e}$ by means of a gauge transformation.  On each sector there is a BRST symmetry that transforms $A_\mu$, $q_I$, auxiliary fields, and ghosts. The latter can be grouped as $b_I , c_I$, needed to exponentiate the constraint and determinant in Eq. \eqref{ident}, and $b,c$, originated from the pure modulus condition in Eq. \eqref{pmod}. 
 The BRST symmetry has a sector-independent algebraic structure that cannot be extended globally, due to specific regularity conditions in each sector \cite{OS-det}.  This is a welcome property as  each BRST can be used to show that partial contributions to observables do not depend on gauge parameters, but not to conclude that the asymptotic space of states is formed by gluons.

\section{Magnetic defects and non-Abelian degrees of freedom}  
\label{nonAdof}

Configurations $\mathscr{A}_\mu \in 
\mathscr{V}(S)$ are created on top of perturbative (topologically) trivial ones, $P_\mu \in \mathscr{V}(I)$,  by means of a singular transformation \cite{tHooft:1977nqb}
 \begin{gather} 
 {\rm Ad}(\mathscr{A}_\mu )  = R(S) {\rm Ad}(P_\mu) R(S)^{-1} +i\,  R(S) \partial_\mu R(S)^{-1} =    R(S) {\rm Ad}(P_\mu -Z_\mu) R(S)^{-1} \;, 
\label{ansat}    \\ 
  R(S) = {\rm Ad}(S)   \makebox[.5in]{,}  {\rm Ad}(Z_\mu) = 
i R(S)^{-1}\partial_\mu R(S) \;.  
\end{gather} 
The use of the adjoint representation  ${\rm Ad}(S)$ eliminates unphysical terms, localized on three-volumes, that would be present when computing 
\begin{equation} 
S P_\mu S^{-1} +i\,  S \partial_\mu S^{-1}   \;.
\label{St}
\end{equation} 
An equivalent procedure to get rid of these terms was introduced in Ref. \cite{reinhardt-engelhardt}. 
Besides the usual covariant field strength $F_{\mu \nu} (\mathscr{A})$, 
it will be useful to define achromoelectric gauge-invariant object 
\begin{equation}
G_{\mu \nu}(\mathscr{A} ) = S^{-1} F_{\mu \nu}(\mathscr{A}) S  
 \makebox[.5in]{,} ( G_{\mu \nu}(\mathscr{A} ), T_\mathscr{A} )  = (  F_{\mu \nu}(A) , n_A )  \makebox[.5in]{,} n_A = S T_A S^{-1}  \;.
\label{ginva}
\end{equation}
Magnetic defects are manifested in field strengths through the commutators of ordinary derivatives $[\partial_\mu, \partial_\nu]$, which are nontrivial when applied on singular mappings,
\begin{gather}
 F_{\mu \nu}(\mathscr{A}) 
= S  \big(  F_{\mu \nu}(P) -F_{\mu \nu}(Z) \big)  S^{-1} \makebox[.5in]{,}
 G_{\mu \nu}(\mathscr{A})  = F_{\mu \nu}(P)  -  F_{\mu \nu}(Z)  \;,
 \label{FPZE}   \\  
 {\rm Ad} (  F_{\mu \nu}(Z) )= i\, 
R(S)^{-1}[\partial_\mu,\partial_\nu ] R(S) \;.
\label{efeZ}
\end{gather}

\subsection{Chains}
\label{ch-sec} 

Let us briefly review some examples. A center vortex worldsheet $\Sigma$ can be created by
\begin{equation}
\bar{S} = e^{i\chi\, \vec{\beta} \cdot \vec{T}} \makebox[.5in]{,} \vec{\beta} \cdot \vec{T} \equiv \vec{\beta}|_q T_q  \makebox[.5in]{,} \vec{\beta} = 2N \vec{w}\;,
\label{cenvor}
\end{equation}
where $\chi$, $\partial^2\chi =0$, is a multivalued phase that changes  by $2\pi$ when going around a path linking $ \Sigma$, and $T_q$, $q= 1, \dots, N-1$, are the Cartan generators.   
The magnetic weight $\vec{\beta}$  is $2N$ times a weight $\vec{w}$ of $\mathfrak{su}(N)$, see Eq. \eqref{weight-vector}. The simplest case corresponds to the fundamental representation $\vec{w} =\vec{w}_i$, $i=1, \dots, N$  \cite{konishi-spanu}, which will be considered from now on.  
In chains, pairs of center vortex branches are matched by monopoles \cite{Reinhardt:2001kf}, \cite{lucho},   
\cite{conf-qg}. In this case, we can write
\begin{equation}
\bar{S} = e^{i\chi\, \vec{\beta} \cdot \vec{T}}\, W \;,
\label{ele-comp} 
\end{equation}   
where the single-valued $W$ creates a closed monopole worldline $\mathcal{C}_{\rm m}$ on $\Sigma$. 
For example,  a pair of 
semi-infinite center vortices is created by using $\chi =\varphi $, $W = 
e^{i\theta \,\sqrt{N} T_{\alpha}} $, where $\varphi$, $\theta$ are polar angles centered at the monopole, and $T_\alpha$, labeled by the adjoint weight (root) $\vec{\alpha} =  \vec{w}- \vec{w}'$, is a combination of root vectors   (cf. Eq. \ref{Tes}). Since the map $W(\pi)$ is a Weyl transformation, 
\begin{equation}
W(\pi)^{-1}\vec{\beta} \cdot \vec{T} \, W(\pi) = \vec{\beta}' \cdot \vec{T}  \;,
 \label{cv-map}    
\end{equation}    
 it interpolates  between two different behaviors, $\bar{S} \sim e^{i\varphi\, \vec{\beta} \cdot \vec{T}}$ and $\bar{S} \sim e^{i\pi \,\sqrt{N} T_{\alpha}}  \, e^{i\varphi\, \vec{\beta}'  \cdot \vec{T}}$, around $\theta = 0$ and $\theta = \pi$, respectively. The factor $e^{i\pi \,\sqrt{N} T_{\alpha}} $ has no effect on gauge-invariant quantities, so that the branches are along $\vec{\beta} \cdot \vec{T}$ and $\vec{\beta}'\cdot \vec{T}$.   
Indeed, the contribution to $G_{\mu \nu}$ is   \cite{reinhardt-engelhardt},
\cite{Reinhardt:2001kf}
\begin{gather} 
 -{\mathcal F}_{\mu \nu}(\bar{Z})     =   2\pi \, \vec{\beta} \cdot \vec{T}  \, \int d^2 \sigma_{\mu \nu}\, \delta^{(4)} (x-y(\sigma_1,\sigma_2)) 
+  2\pi \, \vec{\beta}'  \cdot \vec{T} \, \int d^2 \sigma_{\mu \nu}\, \delta^{(4)} (x- y'(\sigma_1,\sigma_2))  \nonumber \\
  {\cal F}_{\mu \nu}(\bar{Z}) =\frac{1}{2}\, \epsilon_{\mu \nu \rho \sigma} F_{\rho \sigma} (\bar{Z})  \makebox[.5in]{,}  d^2 \sigma_{\mu \nu} = d\sigma_1 d\sigma_2\, \left(\frac{\partial y_{\mu}}{\partial \sigma_1} \frac{\partial y_{\nu}}{\partial \sigma_2} -\frac{\partial y_{\mu}}{\partial \sigma_2} \frac{\partial y_{\nu}}{\partial \sigma_1}\right)\;,
\end{gather}
where the integrals are done over branches with common border at $\mathcal{C}_{\rm m}$ and whose union is $\Sigma$.

\subsection{Chains with monopole fusion}
\label{chain-f}

In order to discuss possible monopole matchings,  let us consider a simple example for $N \geq 3$. At a given time $t$, on a section $\mathbb{R}^3$ of the 4d Euclidean spacetime, let three points be placed on a line at positions $\mathbf{x}_A$, $\mathbf{x}_B$, $\mathbf{x}_C$ (in that order). The map
\begin{equation}
\bar{S} =  e^{i\varphi \vec{\beta}_1\cdot \vec{T}} W(\gamma, \gamma') \makebox[.5in]{,} W(\gamma, \gamma') =W_{12}(\gamma) W_{13}(\gamma')   \makebox[.5in]{,}  W_{ij}(\theta)=e^{i\theta \,\sqrt{N} T_{\alpha_{ij}}} \;, 
\label{threem}   
\end{equation}
where $\gamma $ (resp. $\gamma'$) is the angle that $\mathbf{x}_A$, $\mathbf{x}_B$ (resp. $\mathbf{x}_B$, $\mathbf{x}_C$)  subtend from the observation point $\mathbf{x}$, describes three monopoles joined by center vortices. In effect,  close to the line, to the left of $\mathbf{x}_A$ and to the right of $\mathbf{x}_C$, $\gamma$ and $\gamma'$ tend to zero, i.e., $\bar{S} \sim e^{i\varphi \vec{\beta}_1\cdot \vec{T}}$. The same behavior is verified away from the three points.  
When the segments between $\mathbf{x}_A$, $\mathbf{x}_B$ ($\gamma \to \pi$, $\gamma' \to 0$)
and between $\mathbf{x}_B$, $\mathbf{x}_C$ ($\gamma \to 0$, $\gamma' \to \pi$) are approached, we obtain
\[ 
\bar{S} \sim   e^{i\varphi \vec{\beta}_1\cdot \vec{T}} W_{12}(\pi) = W_{12}(\pi) \,
 e^{i\varphi \vec{\beta}_2\cdot \vec{T}} \makebox[.7in]{\rm and} 
 \bar{S} \sim   e^{i\varphi \vec{\beta}_1\cdot \vec{T}} W_{13}(\pi) = W_{13}(\pi) \,
 e^{i\varphi \vec{\beta}_3\cdot \vec{T}}  \;.  
\] 
 Hence, $\bar{S}$ describes center vortex worldsurfaces meeting at three worldlines  
$\mathbf{x}_{A }(t)$, $\mathbf{x}_{B }(t)$, $\mathbf{x}_{C }(t)$, with common endpoints, that carry adjoint weights   
\begin{equation}
\vec{\delta}_{1}=\vec{w}_1 - \vec{w}_2   \makebox[.5in]{,}
\vec{\delta}_{2} =\vec{w}_2 - \vec{w}_3   \makebox[.5in]{,}
\vec{\delta}_{3} =\vec{w}_3 - \vec{w}_1   \;. 
\label{delta-w}
\end{equation} 
 This array describes a creation-annihilation process with the fusion rule $\vec{\delta}_{1} + \vec{\delta}_{2} + \vec{\delta}_{3} =0$. Four monopole worldlines can be fused in a similar way.  In the general case, the field tensor is a sum over open-surface contributions 
\[ 
-\mathcal{F }_{\mu \nu}(\bar{Z})    = 2\pi \, \sum_{j}\vec{\beta}_j \cdot \vec{T} \, \int d^2 \sigma^j_{\mu \nu}\, \delta^{(4)} (x- y_j(\sigma_1,\sigma_2)) \;.
\]

\subsection{Non-Abelian d.o.f.}  
\label{nonad}

Consider a label $S_0$ in the gauge-fixed partial contribution $Z_{YM}^{(S_0)} $. The left action $S_0 \to U_{\rm e}  S_0 $ simply corresponds to a chromoelectric gauge transformation. On the other hand, the right action
$S_0 \to S_0\, \tilde{U}^{-1} $ generally leads to a new  class $ [S_0\,  \tilde{U}^{-1}] \neq [S_0] $.   
Of course, starting with perturbative configurations ($S_0=I$) no new class is generated,  $ [\tilde{U}^{-1}] = [I] $. In the other cases,  the transformed labels represent a continuum of different
sectors  $\mathscr{V}(S_0\,  \tilde{U}^{-1})$ modulo the equivalence relation in Eq. \eqref{i-rel}. Now, as $\tilde{U}$ is regular, it cannot change the number nor the location of magnetic defects. Then, for each possible distribution of defects, there is a continuum of partial contributions. 
This leads to an important observation: defects possess {\it physical} non-Abelian degrees of freedom. Their relevance can be attributed  to the fact that
$F_{\mu \nu}(Z)$, the second term in the chromoelectric gauge-invariant tensor $G_{\mu \nu}$,  is generally modified. For the above-mentioned examples, we have
\begin{gather}
F_{\mu \nu}(Z) =  \tilde{U} F_{\mu \nu}(\bar{Z} )\, \tilde{U}^{-1} \makebox[.5in]{,} S = \bar{S}\, \tilde{U}^{-1} \;.
\label{rota}  \\
  D_\mu (\tilde{L} )=  \partial_\mu -i[\tilde{L}_\mu, ~]   \makebox[.5in]{,} \tilde{L}_\mu\, = i\, \tilde{U}\,  \partial_\mu  \tilde{U}^{-1}  \;.
\label{Z-comb} 
\end{gather}    
Thus, for a chain and the example with fusion, the monopole currents are,  respectively,   
 \begin{eqnarray} 
&& -D_\nu(\tilde{L})\, {\cal F}_{\mu \nu}(Z) =  2\pi\, 2N\, \tilde{U}\, \vec{\alpha} \cdot \vec{T} \, \tilde{U}^{-1}\, \oint_{\mathcal{C}_{\rm m}}  dy_\mu\, \delta^{(4)}(x-y) \;,
\label{newprop}  \\   
&&   - D_\nu(\tilde{L})\, \mathcal{F }_{\mu \nu}(Z) = 2\pi\, 2N\, \sum_j \tilde{U}\, \vec{\delta}_j \cdot \vec{T}   \, \tilde{U}^{-1}\,  \int_{\gamma_j}  dy^j_\mu\, \delta^{(4)}(x-y^j)  \makebox[.3in]{,}  \sum_j \vec{\delta}_j =0 \;,
\label{fusion-m} 
\end{eqnarray}       
 which are covariantly conserved.  
Note also that the second term in the usual field strength continues to be 
along the Cartan sector, $  S F_{\mu \nu}(Z) \, S^{-1} =  {\cal F}_{\mu \nu}(\bar{Z})$ (cf. Eqs. \eqref{FPZE},  \eqref{efeZ}).

\section{Gaussian gauge-invariant smoothing} 
\label{simpli}
 
The  Wilson loop for quarks in an irreducible $\mathscr{D}$-dimensional representation ${\rm D}$ is
\begin{eqnarray} 
{\mathcal W}_{\rm e}[\mathscr{A}]  =  \frac{1}{\mathscr{D}} \, {\rm tr}\, {\rm D}  \left( P \left\{ e^{i \oint_{{\cal C}_{\rm e}} dx_\mu\,  A_{\mu}(x)  } \right\} \right) \;.
\label{Wloop}  
\end{eqnarray}   
When thin configurations are considered, i.e., $P_\mu = 0$ in Eqs. \eqref{ansat}, \eqref{St}, the result for a chain coincides with that for a center vortex placed at the same location. The non-Abelian d.o.f. do not play a role either. Indeed, the Wilson loop is given by 
\begin{gather} 
\mathfrak{z}({\cal C}_{\rm e})   =  \frac{1}{\mathscr{D}}\, {\rm tr}\, {\rm D} (S_{\rm f} \,  S_{\rm i}^{-1})    \;,
\label{wloo}  \\ 
S = e^{i\chi\, \vec{\beta} \cdot \vec{T}} \, W \,  \tilde{U}^{-1} \makebox[.5in]{,}  S_{ f} \, S_{ i}^{-1} = e^{i(\chi_f -\chi_i)\, \vec{\beta} \cdot \vec{T}} = e^{i 2\pi \, \vec{\beta} 
\cdot \vec{T}\, L({\cal C}_{\rm e}, \Sigma)}    \;.
 \label{ini-fin}     
\end{gather}    
This only depends on the linking number $L({\cal C}_{\rm e},  \Sigma)$ between ${\cal C}_{\rm e}$ and $\Sigma$. However, the ensemble measure would in principle be generated by path-integrating general field fluctuations $P_\mu$ around magnetic defects, which  might differentiate between  center vortices and chains. Answering if this is the case in the YM context is a difficult task. Instead, in this section we shall discuss a simple example to get some insight about the possible effects. 
 
\subsection{Simple Gaussian model} 
 
For  general configurations in $\mathscr{V}(S)$ (cf.  Eqs. \eqref{ansat}, \eqref{St}) we have 
\begin{eqnarray}
 {\mathcal W}_{\rm e}[\mathscr{A}] 
&=& {\mathcal W}_{\rm e}[P]  \; \mathfrak{z}({\cal C}_{\rm e})  \;. 
\end{eqnarray}  
The linking number can be equated to the intersection number $I(S({\cal C}_{\rm e}), \Sigma)$ between $S({\cal C}_{\rm e})$ (a surface whose border is ${\cal C}_{\rm e}$) and $ \Sigma$, 
\begin{gather}
I(S({\cal C}_{\rm e}), \Sigma)= \frac{1}{2} \int d^2 \tilde{\sigma}_{\mu \nu} \int d^2 \sigma_{\mu \nu}\,
\delta^{(4)} (w(s,\tau)- y(\sigma_1,\sigma_2))\;, \\
d^2 \tilde{\sigma}_{\mu \nu} = \frac{1}{2} \epsilon_{\mu \nu \alpha \beta}\,  d\tau\,  ds\, \left(\frac{\partial w_{\alpha}}{\partial \tau} \frac{\partial w_{\beta}}{\partial s} -\frac{\partial w_{\alpha}}{\partial s} \frac{\partial w_{\beta}}{\partial \tau}\right)\;, 
\end{gather}  
where $w(s,\tau)$ is a parametrization of $S({\cal C}_{\rm e})$ \cite{reinhardt-engelhardt}. Since both vortex orientations will be taken into account, the antifundamental  weights can be disregarded.  The topological contribution may be written in terms of ${\mathcal F}_{\mu \nu}(Z)$ as  follows: i) consider a general configuration $\bar{S}$ with defects  such that ${\mathcal F}_{\mu \nu}(\bar{Z})$ is along the Cartan sector; ii) note that if a chain links ${\mathcal C}_{\rm e}$ then one of the associated vortex branches crosses  $S({\cal C}_{\rm e})$; and iii) use that any magnetic weight $\vec{\beta}_i$ satisfies  
\begin{equation} 
   {\rm D} \big( e^{i 2\pi\, \vec{\beta}_i \cdot \vec{T}} \big) =  {\rm D} \big( e^{-  i \frac{i2\pi}{N} } I \big) = e^{i2\pi\, \vec{\beta}_i \cdot \vec{w}_{\rm e} } \, I_{\mathscr D} \makebox[.5in]{,} \mathfrak{z}({\cal C}_{\rm e}) = \big( e^{i 2\pi\, \vec{\beta}_i \cdot \vec{w}_{\rm e} } \big)^{I(S({\cal C}_{\rm e}), \Sigma) }  \;,
 \label{cen} 
\end{equation}
where the tuple $\vec{w}_{\rm e}$ is any weight of the quark representation (we can choose the highest) and $I_{\mathscr D}$ is the ${\mathscr D} \times {\mathscr D}$ identity matrix.
Therefore, we can write
\begin{equation}
\mathfrak{z}({\cal C}_{\rm e}) =  e^{-\frac{i}{2} \int d^4x\, \big(  s_{\mu \nu} \tilde{U} \vec{w}_{\rm e}\cdot \vec{T } \tilde{U}^{-1} , \; {\mathcal F}_{\mu \nu}(Z) \big)} \makebox[.5in]{,} s_{\mu \nu}(x) = \int_{S({\cal C}_{\rm e})} d^2 \tilde{\sigma}_{\mu \nu} \, \delta^{(4)}(x- w(s,\tau))\;,
\label{IandII-Z}
\end{equation}  
where $\tilde{U}(x)$ is any regular single-valued configuration defined on $\mathbb{R}^4$. 
From Eq. \eqref{rota}, this quantity is $\tilde{U}$-independent, and using i)-iii) we recover Eq.  
\eqref{cen} as long  as the monopoles do not touch $S({\cal C}_{\rm e})$. 
 In addition, the Petrov-Diakonov representation of the Wilson loop \cite{PD} may be used to
 rewrite  the effect of fluctuations as an integral over periodic paths,    
\begin{eqnarray}    
&& {\mathcal W}_{\rm e}[P]   \propto \int [dg]_{\rm P}\, e^{ i\oint dx_{\mu}\,  \big( g^{-1} P_\mu\, g 
+i\, g^{-1 }  \partial_\mu  g , \, \vec{w}_{\rm e}|_q T_q \big)  } \;.
\label{wna} 
\end{eqnarray}  
 For completeness, and to settle notation and conventions, group-coherent states and the path-integral representation of holonomies are briefly reviewed in Appendix \ref{gcoh-s}.  After extending the paths to $\tilde{U}(x) \mid g(s) = \tilde{U}(x(s)) $, we can apply  Stokes' theorem and join $ {\mathcal W}_{\rm e}[P]$ with the center element in  Eq. \eqref{IandII-Z}, thus obtaining 
\begin{gather} 
 {\mathcal W}_{\rm e}[\mathscr{A}] 
   \propto \int [\mathcal{D}\tilde{U}]\,  e^{ \frac{i}{2}\int d^4x \,    \left(  {\mathcal Y}_{\mu \nu} (P, \tilde{U})   -
{\cal F}_{\mu \nu}(Z) \, ,\,  s_{\mu \nu} \tilde{U}   \vec{w}_{\rm e} \cdot \vec{T} \, \tilde{U}^\dagger  \right)  }  \;,
\label{represe}  \\  
 {\mathcal Y}_{\mu \nu}  = \frac{1}{2}\, \epsilon_{\mu \nu \rho \sigma}  Y_{\rho \sigma} \makebox[.5in]{,}
 Y_{\mu \nu}(P,\tilde{U}) = D_\mu(\tilde{L})\,(P_\nu - \tilde{L}_\nu)  - D_\nu(\tilde{L})\, (P_\mu -\tilde{L}_\mu)  \;,
\label{dual-t}  
\end{gather}
where $\tilde{L}_\mu$  was defined in Eq. \eqref{Z-comb}, and
\begin{eqnarray}
 \partial_\mu P_\nu^{\tilde{U}^{\dagger}} -  \partial_\nu P_\mu^{\tilde{U}^{\dagger}} 
= \tilde{U}^\dagger  Y_{\mu \nu} (P, \tilde{U})\, \tilde{U}  \;.
\label{propga}
\end{eqnarray}
Now, let us replace the observable in Eq. \eqref{represe} by the smoothed variable
\begin{eqnarray}
 \mathsf{W} [P,\bar{Z}] = \lefteqn{ \int [\mathcal{D}\tilde{U}]\, e^{-\int d^4x\,  \frac{1}{4g_{\rm e}}  \left( \mathcal{Y}_{\mu \nu} (P, \tilde{U})   -  \mathcal{F}_{\mu \nu}(Z) \right)^2 -\int d^4x\, \frac{\eta^2}{2} (P_\mu -\tilde{L}_\mu )^2}  } \nonumber \\
 && \times \, \,  e^{ \frac{i}{2}\int d^4x \,    \left(  {\mathcal Y}_{\mu \nu} (P, \tilde{U})   -
{\cal F}_{\mu \nu}(Z) \, , \,  s_{\mu \nu} \tilde{U}   \vec{w}_{\rm e} \cdot \vec{T} \, \tilde{U}^\dagger  \right)  }   \;,
\label{quad} 
\end{eqnarray} 
where we have included a gauge-invariant mass term.

\subsection{Dual representation} 
\label{dual-lambda}

 In order to obtain a single valued  $\mathscr{A}_\mu$ in Eq. \eqref{ansat},  the components of $P_\mu$ rotated by $S$ should vanish at the defects. Accordingly, the path-integral over $P_\mu$ has to be performed imposing these regularity conditions.  
Still, it is possible to integrate $ \mathsf{W} [P,\bar{Z}]$ in Eq. \eqref{quad} without them, and relate both results by means of a factor $R [\bar{Z}]$,
\[
\int [{\mathcal D}  P_\mu]_{\rm r.c.}\,  \mathsf{W} [P,\bar{Z}] = R [\bar{Z}] \int [{\mathcal D}  P_\mu]\,  \mathsf{W} [P,\bar{Z}] \;.
\]
The ratio $R [\bar{Z}]$ contains information about the distribution of center vortices,  providing their intrinsic properties. 
Introducing a Lie algebra-valued tensor $\Lambda_{\mu \nu}$, we can rewrite
\begin{eqnarray} 
&&  \int [{\mathcal D}  P_\mu]\,  \mathsf{W} [P,\bar{Z}] = \int  [\mathcal{D}\Lambda_{\mu \nu}][\mathcal{D}\tilde{U}]\, e^{-\frac{1}{8\eta^2} \, \frac{1}{(4\pi N)^2}  \int d^4x\,  \left( \Phi_\mu , \,  \Phi_\mu    \right)  } \nonumber \\
&& \times \, \,  e^{-\int d^4x\,    \frac{1}{4g^2} \,  \left(  \Lambda_{\mu \nu} - 2\pi   s_{\mu \nu} \tilde{U} \vec{\beta}_{\rm e} \cdot \vec{T} \, \tilde{U}^\dagger\right)^2  } \,   e^{-\frac{i}{2} \, \frac{1}{4\pi N}  \int d^4x\,  \left( \Lambda_{\mu \nu} , \,  
{\cal F}_{\mu \nu}(Z) \right)  }  \makebox[.5in]{,} g= 4\pi N/g_{\rm e} \;, \nonumber 
 \end{eqnarray}
where $\Phi_\mu =  \epsilon_{\mu \nu \rho \sigma  }  D_\nu (\tilde{L})\,  \Lambda_{\rho \sigma} $ and $\vec{\beta}_{\rm e} = 2N \vec{w}_{\rm e} $. 
Then, using a gauged version of the usual Hodge 
decomposition,
\begin{equation} 
  \Lambda_{\mu \nu} = Y_{\mu \nu} + B_{\mu \nu} \makebox[.5in]{,} Y_{\mu \nu}(\Lambda,\tilde{U} ) = D_\mu(\tilde{L})\, (\Lambda_\nu -\tilde{L}_\nu) 
 - D_\nu(\tilde{L}) \,(\Lambda_\mu -\tilde{L}_\mu)    \;,
\end{equation}  
$ D_\nu(\tilde{L}) \, B_{\mu \nu} = 0$, we see that $B_{\mu \nu}$ couples with the curl of $s_{\mu \nu}$, which is localized on ${\cal C}_{\rm e}$. Also note that the limit $\eta \to 0$ enforces the constraint $\Phi_\mu = 0 $, whose solution is
\begin{equation} 
  \Lambda_{\mu \nu} = Y_{\mu \nu}(\Lambda,\tilde{U} )  \;.
\end{equation} 
In this respect,  $\tilde{L}_\mu $ defined in Eq. \eqref{Z-comb} has the form of a pure gauge, so that\footnote{The general theory to deal with this type of non-Abelian Hodge decomposition, with zero curvature connections, was developed in Ref. \cite{Oxf}. }  
\[
\epsilon_{\sigma \rho \mu \nu  } D_\rho(\tilde{L})\,   D_\mu(\tilde{L}) \, X_\nu = (1/2)\,  \epsilon_{\sigma \rho \mu \nu  } [D_\rho(\tilde{L}),    D_\mu(\tilde{L})] \, X_\nu = 0 \;. 
\] 
Thus, for small $\eta$, the Gaussian smoothing leads to   
\begin{gather} 
\int [{\mathcal D}  P_\mu]_{\rm r.c.}\,  \mathsf{W} [P,\bar{Z}] \approx  R [\bar{Z}] \int [\mathcal{D}\Lambda_{\mu }]\,  \mathsf{V}[\Lambda,\bar{Z}]  \;,
 \label{duality} \\    
  \mathsf{V} [\Lambda,\bar{Z}]  =   \int [\mathcal{D}\tilde{U}] \; e^{ -\int d^4x\,    \frac{1}{4g^2} \,   \left(Y_{\mu \nu}(\Lambda,\tilde{U} ) - 2\pi s_{\mu \nu} \tilde{U}  \vec{\beta}_{\rm e} \cdot \vec{T} \,  \tilde{U}^\dagger \right)^2} \, e^{-\frac{i}{4\pi N}    \int d^4x \,     \big( \Lambda_\mu -\tilde{L}_\mu \, , D_\nu(\tilde{L})\, {\cal F}_{\mu \nu}(Z) \big) }  \;.
\label{Vrep}
\end{gather}    
In particular, for a given distribution of monopole loops, using Eqs. \eqref{newprop} and  
 \begin{eqnarray} 
    \big( \Lambda_\mu -\tilde{L}_\mu \, , \,\tilde{U}\, (\vec{\alpha}\cdot \vec{T})\, \tilde{U}^{-1}  \big)   = 
   \big(\tilde{U}^{-1} \Lambda_\mu\, \tilde{U} 
+i\, \tilde{U}^{-1 }  \partial_\mu  \tilde{U} , \vec{\alpha}\cdot \vec{T}\big)   \;, 
 \end{eqnarray}
 we obtain
\begin{eqnarray}
 \mathsf{V} [\Lambda,\bar{Z}] = \lefteqn{  \int [\mathcal{D}\tilde{U}] \; e^{ -\int d^4x\,    \frac{1}{4g^2} \,   \left(Y_{\mu \nu}(\Lambda,\tilde{U} )- 2\pi s_{\mu \nu} \tilde{U}  \vec{\beta}_{\rm e} \cdot \vec{T} \,  \tilde{U}^\dagger \right)^2} }  \nonumber \\ 
&& \times \, \,  e^{  i  \sum_{k}  \oint_{\mathcal{C}_k}    dx^{(k)}_\mu \,    \big( \tilde{U}^{-1} \Lambda_\mu\, \tilde{U} 
+i\, \tilde{U}^{-1 }  \partial_\mu  \tilde{U} , \vec{\alpha}\cdot \vec{T} \big)_k } \;. 
\label{VL}
\end{eqnarray}  
The index $k$ in the parenthesis means that $\tilde{U}(x)$ is to be taken as 
$\tilde{U}(x^{(k)}(s_k))$. Likewise, monopole fusion (cf. Eq. \eqref{fusion-m}) implies additional factors involving
\begin{eqnarray}  
 e^{   i  \sum_{j=1}^n \int_{\gamma_j} dx_\mu\,   \big( \tilde{U}^{-1} \Lambda_\mu\, \tilde{U}  
+i\, \tilde{U}^{-1 }  \partial_\mu  \tilde{U} , \, \vec{\delta} \cdot \vec{T}\big)_j } \makebox[.5in]{,} 
\sum_j \vec{\delta}_j =0 \;.
\label{nmon}   
\end{eqnarray}

Under non-Abelian magnetic gauge transformations 
\begin{equation} 
\Lambda_\mu \to U_{\rm m}\,  \Lambda_\mu U_{\rm m}^{-1} +i\, U_{\rm m}\, \partial_\mu U_{\rm m}^{-1}   \;,
\end{equation} 
$ \mathsf{V} [\Lambda,\bar{Z}]$ is in principle invariant, as this change can be absorved by $ \tilde{U} \to U_{\rm m}\, \tilde{U} $. Note that in Eq. \eqref{VL}, the non-Abelian d.o.f. $\tilde{U}$ are coupled to $\Lambda_\mu$ on the whole spacetime, which 
hinders the integration over the group. Yet we can get some insights from the formal expressions.
In the dual representation of the Wilson loop average, the effect of linking numbers is encoded as a frustration $2\pi s_{\mu \nu}  \vec{\beta}_{\rm e} \cdot \vec{T} $ in the action for the dual gauge field $\Lambda_\mu$. Moreover, monopoles with non-Abelian d.o.f. $\tilde{U}$ and fusion rules have a nontrivial indirect effect through the coupling to a dual non-Abelian gauge field $\Lambda_\mu$.

\section{Percolating center vortices and chains}  
\label{first}   

In this section we shall propose a possible measure to compute center-element averages in 4d ensembles of percolating center vortices and chains.   
For this aim, let us recall the situation in 3d Euclidean spacetime, where the confining and  
 deconfining phases can be described by an effective complex vortex field $V$.
  
 \subsection{Percolating center vortices in 3d} 
 \label{v3d}
 
 In three dimensions, the  average of the fundamental Wilson loop over an ensemble of  center vortices, with small (positive) stiffness $\frac{1}{\kappa}$  and repulsive contact   interactions, is represented by \cite{Oxman-Reinhardt-2017} 
\begin{gather}
\frac{Z_{\rm v}^{(3)} [s_\mu]}{Z_{\rm v}^{(3)} [0]} \makebox[.5in]{,} Z_{\rm v}^{(3)} [s_\mu]
=\int [{\cal D}V][{\cal D}\bar{V}]\,   e^{-\int
	d^{3}x\,\left[ \frac{1}{3 \kappa}\,  \overline{D_\mu V} D_\mu V +\frac{1}{2\zeta}\, (\overline{V} V -v^2)^2  \right]} \;,
\label{mix}  \\
D_\mu = \partial_{\mu}- i \frac{2\pi }{N} s_\mu  \makebox[.5in]{,} s_\mu =  \int_{S({\cal C}_{\rm e})} d\tilde{\sigma}_\mu \, \delta^{(3)}(x-w(s, \tau))  \;.
\end{gather}
 In the normal phase ($v^2 < 0$), as the vacuum is at $V=0$, we have to deal with the complete complex field $V$. 
In the percolating phase ($v^2 >  0$), the computation of the Wilson loop is a hard problem due to the large quantum fluctuations of the Goldstone modes $\gamma(x)$, $V(x) = \rho(x)\, e^{i\gamma(x)}$. In order to discuss this case, we kept the soft degrees of freedom $V=v\, e^{i\gamma}$,  
\begin{eqnarray} 
Z_{\rm v}^{(3)} [s_\mu]
\approx \int [{\cal D}\gamma] \,   e^{-\int
	d^{3}x\,  \frac{v^2}{3\kappa}\, D_\mu (e^{-i\gamma}) D_\mu (e^{i\gamma}) } \;,
	\label{G3-latt}
\end{eqnarray}
and  switched to the lattice, where the finite spacing  takes care of possible phase singularities in $\gamma$. This amounts to considering the frustrated 3d XY model
\begin{equation}
\beta \, \sum_{x, \mu }  \mathrm{Re} \left[ 1- 
 e^{i \gamma(x + \hat{\mu})} e^{-i \gamma(x)} e^{-i \alpha_\mu ({x})}   \right] 
  \;,
  \label{disc-3d}
\end{equation}
 where the frustration $ \alpha_\mu (\mathbf{x})$ takes the value $ \frac{2\pi}{N}$, if the surface $S
 ({\cal C}_{\rm e})$ is crossed by the link (in the direction of the normal to $S({\cal C}_{\rm e})$), and is zero otherwise. For a discussion in the context of superfluids, see Ref.  \cite{kleinert1989gauge}. As is well-known, the continuum limit is attained at   
 $\beta_{\rm c}  \approx 0.454 $, where a Wilson loop area law with $N$-ality is obtained as an extensive property of the ensemble \cite{Oxman-Reinhardt-2017}.  
Summarizing,  while closed worldlines are naturally described by a complex field $V$, in a  condensate their description can be approximated by a different object, namely, a compact real field $\gamma$ representing the Goldstone modes. 

\subsection{Percolating center vortices in 4d} 

In four dimensions, the effective description for a general ensemble of worldsurfaces would be in terms of a second quantized string field, however, percolating worldsurfaces may also be approximated by a simpler object. Indeed, in Ref. \cite{Rey}, a condensate was described by an action functional for a complex string Higgs field with frozen modulus. The Goldstone modes can be read in the phase $\gamma$ of this field, but this time it is defined on strings rather than at spacetime points. In that work, the different possibilities were parametrized in terms of an Abelian gauge field $\Lambda_\mu$ such that the phase $\gamma_{\Lambda}$ is the line integral of $\Lambda_\mu$ along the string. Moreover, the lattice partition function of the string field model  was approximated by a field model
\begin{equation}
\beta \,\sum_{\mathbf{x}, \mu < \nu } \mathrm{Re} \left[ 1- V_\mu(x)  V_\nu(x + \hat{\mu}) V^{-1}_\mu(x + \hat{\nu}) V^{-1}_\nu(x) \, e^{i a^2 B_{\mu \nu}(x)  }  \right] \makebox[.5in]{,} V_\mu(x) = e^{ia \Lambda_\mu(x)}  \;, 
\end{equation}
where  $B_{\mu \nu}$ is an external smooth Kalb-Ramond field coupled to the worldsurfaces.\footnote{In Ref. \cite{Rey}, the action has an additional term that gives a dynamics to $B_{\mu\nu}$.} In other words, the Goldstone modes for Abelian condensates of closed surfaces are represented by Abelian gauge fields (for related ideas, see Refs. \cite{Julia1979}-\cite{Gri2010}). This observation, together with: 
\begin{itemize}
 
\item the previous discussion about 3d center vortex condensates with frustration

\item the presence of non-Abelian d.o.f. on center vortices (see section \ref{nonad})

\item the natural generation of dual non-Abelian gauge fields $\Lambda_\mu$ with frustration when representing some Wilson loop averages (see section \ref{dual-lambda})

\end{itemize}
lead us to replace the link variables in Eq. \eqref{disc-3d} by non-Abelian ones $V_\mu \in SU(N)$ and propose
\begin{gather}
\frac{Z_{\rm v}^{\rm latt} [\alpha_{\mu \nu}]}{Z_{\rm v}^{\rm latt} [0]} \makebox[.5in]{,} Z_{\rm v}^{\rm latt} [\alpha_{\mu \nu}] = \int [{\cal D} V_\mu] \, e^{-\beta  \sum_{\mathbf{x}, \mu < \nu } \mathrm{Re}\;  {\rm tr} [ I -  V_\mu(x) V_\nu(x + \hat{\mu}) V^\dagger_\mu(x + \hat{\nu}) V^\dagger_\nu(x) e^{-i \alpha_{\mu \nu}(x)  }  ]  } \;,
\label{cvlatt} 
\end{gather}
as a measure to average center elements over a 4d ensemble of  percolating center vortices. The frustration $\alpha_{\mu \nu}$ is only  nontrivial on plaquettes $x,\mu,\nu$ that intersect $S(\mathcal{C}_{\rm e})$, where it satisfies
$ e^{-i  \alpha_{\mu \nu} }  = e^{-i2\pi\,  \vec{\beta}_{\rm e} \cdot \vec{T}} $. Now, let us discuss 
the meaning of the definition (\ref{cvlatt}) on its own.
Note that the usual properties of  ordinary integrals over the group  imply that, for an arbitrary order in powers of $\beta$, the contribution to $Z_{\rm latt}^{(4)} [\alpha_{\mu\nu}]$ is originated from plaquettes that form closed surfaces \cite{Creutz}.  Surfaces that link $\mathcal{C}_{\rm e}$ will intersect $S(\cal{C}_{\rm e})$ a number of times, gaining a factor $  e^{\mp i  \alpha_{\mu \nu} } = e^{\pm i2\pi\,  \vec{\beta}  \cdot   \vec{w}_{\rm e} } \, I$  for every  intersection point. In effect, acting with $ e^{\mp i  \alpha_{\mu \nu} }$ on a basis of the fundamental representation formed by weight vectors $|\phi_{w_i} \rangle $, $i=1, \dots, N$, we get $i$-independent quantities $ e^{-i2\pi\,  \vec{\beta}_{\rm e} \cdot \vec{w}_i} = e^{-i2\pi\,  \vec{w}_{\rm e} \cdot \vec{\beta}_i} $, that is, the eigenvalues of  ${\rm D} (e^{\pm  i \frac{i2\pi}{N} } I ) $ (cf. Eq. \eqref{cen}).  
Then,  the fingerprints of center vortices are present in Eq.  \eqref{cvlatt} as it involves the same center elements
$\mathfrak{z}({\cal C}_{\rm e}) $ that are generated when a quark Wilson loop in representation ${\rm D} $ is linked by a center vortex, averaged over an ensemble of plaquettes distributed on closed surfaces. For larger $\beta$ values, larger and multiple closed worldsheets become more important and, as such, performing the average is a hard problem.  Note that in the partition function for the frustrated 3d XY model, we can also conclude that the nontrivial contributions to $ \prod_x \int_{-\pi}^{\pi} d\gamma(x) $ 
are originated from links distributed along closed loops accompanied by a center element. In that case, the difference is that the effective description \eqref{mix}, which includes the normal phase, can be derived from the microscopic one \cite{Oxman-Reinhardt-2017}.
 
\subsection{Percolating chains in 4d}

Relying on the Gaussian smoothing of the Wilson loop, we showed that general fluctuations induce a combined effect of center vortices and monopoles with non-Abelian d.o.f.  (cf. Eqs. \eqref{duality}, \eqref{VL}). Here, we shall consider an effect on center-element averages that distinguish 
between percolating center vortices and chains, which will be included as a phenomenological property that YM ensembles might have. Center vortex branches attached in pairs to fixed closed worldlines ${\cal C}^{\rm latt}_k$ ($k=1,\dots, n$) on the hypercubic lattice can be included  by means of the partial contribution
\begin{gather}
 Z^{\rm latt}_{\rm mix} [s_{\mu \nu}]\big|_{\rm p} \propto   \int [{\cal D} V_\mu] \, e^{-\beta  \sum_{\mathbf{x}, \mu < \nu } \mathrm{Re}\;  {\rm tr} [ I -  V_\mu(x) V_\nu(x + \hat{\mu}) V^\dagger_\mu(x + \hat{\nu}) V^\dagger_\nu(x) e^{-i \alpha_{\mu \nu}(x)  }  ]  } \, {\mathcal W}^{(1)}_{\rm Ad} \dots  {\mathcal W}^{(n)}_{\rm Ad}   \;,
\label{fixmon-latt} \\
{\mathcal W}^{(k)}_{\rm Ad}    =  \frac{1}{N^2-1} \, {\rm tr}\,   \left( \prod_{(x,\mu)\in \,
{\cal C}^{\rm latt}_k } {\rm Ad} \big( V_{\mu}(x) \big)   \right)  \;.   
\label{w-latt} 
\end{gather}  
While the group integral of ${\rm Ad }(V_\mu)|_{AB}$ is trivial (see Appendix \ref{ortog}), the integral of the combination $  {\rm Ad }(V_\mu)|_{AB} \, {V_\mu}|_{ij} \, V_\mu^{\dagger}|_{kl}$ is nontrivial, since $N  \otimes  \bar{N}$ contains an adjoint irrep that can form a singlet with the first factor. Then, the relevant configurations in Eq. \eqref{fixmon-latt} 
occur when the link along the adjoint loops combine with fundamental-antifundamental pairs of variables generated by the Wilson action. In other words, the contribution
to $ Z^{\rm latt}_{\rm mix} [s_{\mu \nu}]\big|_{\rm p}$ derives from plaquettes distributed on open surfaces that meet in pairs at the adjoint loops, 
forming closed two-dimensional arrays, and including disconnected closed parts.  Whenever the surface $S({\cal C}_{\rm e})$ is  intersected, the configuration will be accompanied by a center element. That is, the closed surfaces and arrays can be identified with  center vortices and chains, respectively.     
At small $\beta$, the leading contribution is given by plaquettes distributed on the faces of elementary cubes with edge at ${\cal C}^{\rm latt}_k$ (see Ref. \cite{Creutz}, p. 62).  On the other hand, as $\beta$ is increased, percolating branches with fixed boundaries ${\cal C}^{\rm latt}_k$ are expected to occur. 

\subsection{Ensemble integration of monopole Wilson loops}

In section \ref{frules}, we will include monopole fusion; for now, the average of center elements over the ensemble mixture should sum the partial contributions considering all possible numbers and configurations of monopole loops. However, in the lattice, the calculation of the partition function could only be accessed by computer simulations. Hence, let us switch to the continuum description of the ensemble where Eqs. \eqref{fixmon-latt} and \eqref{w-latt} become
\begin{gather} 
 Z_{\rm mix} [s_{\mu \nu}]\big|_{\rm p} \propto  \int [{\cal D}\Lambda_\mu] \,   e^{-\int d^4x\,    \frac{1}{4g^2} \,   \left( F_{\mu \nu}(\Lambda ) - 2\pi s_{\mu \nu} \vec{\beta}_{\rm e} \cdot \vec{T}  \right)^2  }  {\mathcal W}^{(1)}_{\rm Ad}[\Lambda] \dots  {\mathcal W}^{(n)}_{\rm Ad}[\Lambda]   \;,
\label{fixmon}   \\
{\mathcal W}^{(k)}_{\rm Ad}[\Lambda]  =  \frac{1}{N^2-1} \, {\rm tr}\,  {\rm Ad} \left( P \left\{ e^{i \oint_{{\cal C}_k} dx_\mu\,  \Lambda_{\mu}(x)  } \right\}  \right)  \;.
\end{gather}  
 $F_{\mu \nu} (\Lambda )$ is the usual field strength \eqref{nona-f}, computed for a non-Abelian gauge field $\Lambda_\mu$, and $g$ is a dual coupling  ($\beta \sim  \frac{1}{g^2}$). Accordingly, 
the complete average of center elements turns out to be
\begin{equation}
 \frac{Z_{\rm mix} [s_{\mu \nu}]}{Z_{\rm mix} [0]} 
\makebox[.5in]{,}   Z_{\rm mix} [s_{\mu \nu}] = \int [{\cal D}\Lambda_\mu] \,   e^{-\int d^4x\,    \frac{1}{4g^2} \,   \left( F_{\mu \nu}(\Lambda ) - 2\pi s_{\mu \nu} \vec{\beta}_{\rm e} \cdot \vec{T}  \right)^2  }
Z_{\rm loops} [\Lambda] \;,
\label{funZ}
\end{equation} 
$Z_{\rm loops} [\Lambda] =1 + Z_1 + Z_2 + \dots$ , where $Z_n$ represents a gas of $n$  closed worldlines,
\begin{gather}
 Z_n [\Lambda]= \int [Dm]_n  \prod_{k=1 }^n  \,  e^{- \int_0^{L_k}  ds_k\,  \left[  \frac{1}{2\kappa}\, \dot{u}^{(k)}_\mu 
\dot{u}^{(k)}_\mu +
 \mu \right] } \,  {\mathcal W}^{(k)}_{\rm Ad}[\Lambda]   \;,
 \label{coms}   \\
 u_\mu(s) =\frac{dx_\mu}{ds}  \in S^3  \makebox[.5in]{,}   \dot{u}_\mu(s) =\frac{du_\mu}{ds}  \;.
\end{gather}
The phenomenological dimensional parameters $\mu$ and $1/\kappa$ are associated with tension and stiffness, respectively. These are the simplest properties a monopole loop might have, that is, a weight depending on the loop size and curvature. In the lattice, stiffness represents possible correlations between the link orientations along the loop. In what follows, the parameter $1/\kappa$
will be important to obtain a well-defined continuum limit when the loops are discretized as polymers and thought of as a growing monomer process. At the end, the partition function will assume a simplified form in the region of small but nonzero values of stiffness, where the growth almost behaves as a random walk only weighted by the total size of the loop. 

The measure $[Dm]_n $ implements the integral over paths starting and ending at $x_k$, with tangent vector $u_k$.   
Therefore,  
\begin{eqnarray}
 Z_{\rm loops}[\Lambda] = \sum_n \frac{1}{n!}\, \prod_{k=1 }^n   \int_{0}^{\infty}\;
\frac{dL_{k}}{L_{k}}  \int\; dv_k  \int  [dx^{(k)}]_{v_k,v_k}^{L_k}  \,  e^{- \int_0^{L_k}  ds_k\,  \left[  \frac{1}{2\kappa}\, \dot{u}^{(k)}_\mu 
\dot{u}^{(k)}_\mu +
 \mu \right] } \, {\mathcal W}^{(k)}_{\rm Ad}[\Lambda]   \;,   
 \label{ens-ad}
\end{eqnarray}
where $v$ stands for the pair of variables $x, u$ and $[dx]_{v,v}^{L}$ path-integrates over a closed worldline $x(s)$ with fixed length $L$, starting and ending at $v$. 
Then, the partition function adds up to
\begin{gather} 
Z_{\rm loops} [\Lambda]=   e^{  \int_{0}^{\infty}\frac{dL}{L}\;  \int  dv \,  {\rm tr}\, Q(v, v,L)  }  \;,
\label{pfun}  \\
 Q(x,u,x_0,u_0,L)  =  \int  [dx(s)]_{v,v_0}^L \, e^{- \int_0^{L}  ds\,  \left[  \frac{1}{2\kappa}\, \dot{u}_\mu 
\dot{u}_\mu +
 \mu  \right] } \, {\rm Ad} \left( \Gamma[\Lambda] \right)  
 \label{mat-con}  \;, \\
\Gamma[\Lambda]   =  P \left\{ e^{i \int dx_\mu\,  \Lambda_{\mu}(x)  } \right\} \;,
\end{gather}  
where $\Gamma[\Lambda]$ is the holonomy for an open path $x(s)$ with initial and final conditions $v_0$, $v$. In order to go further, we can follow 
Refs. \cite{GBO}, \cite{ref24}.   
Let us summarize the main steps adapted to the present scenario. As usual, ${\rm Ad}(\Gamma)$ can be associated with an ``evolution'' operator
\begin{equation} 
P \left\{ e^{-\int_0^L ds\, H(s)  } \right\}  \makebox[.5in]{,} H(s)= H(x(s),u(s)) \makebox[.5in]{,} H(x,u)=-i\, u_\mu {\rm Ad} \big(\Lambda_{\mu}(x) \big) \;.
\label{Hg}
\end{equation} 
The path ordering is obtained from the discretized expression
\begin{equation}
P \left\{ e^{-\int_0^L ds\, H(s)}  \right\} \Big|_{\rm d} =  e^{-H(x_M,u_M)\, \Delta L}  \dots\, e^{-H(x_2,u_2)\, \Delta L}  \,   e^{-H(x_1,u_1) \Delta L} \;,  
\end{equation} 
by taking  the $\Delta L \to 0$, $M \to \infty$ limit, with $L = M  \Delta L$. Accordingly, $Q(v,v_0, L)$ is obtained from
\[ 
Q(x, u, x_0, u_0, L) |_{\rm d}  =  Q_{M} (x,u,x_{0}, u_{0})  \makebox[.5in]{,} x=x_M \;, u=u_M \;,
\]  
\begin{eqnarray} 
\lefteqn{ Q_{M}(x_M,u_M,x_0, u_0) =} \nonumber \\
&& =  \int  \prod_{k=1}^{M-1} d^3x_k \, du_k    \prod_{n=1}^M \psi(u_n -u_{n-1}) \,  \delta(x_n - x_{n-1} -
u_n\Delta L)   \nonumber \\
&&  
\times e^{-\sum _{n=1}^M (\mu + \phi(x_n)) \Delta L} \, e^{-H(x_M,u_M)\, \Delta L} \dots \, 
e^{-H(x_2,u_2)\, \Delta L}  \,   e^{-H(x_1,u_1) \Delta L}   \;,
\label{disc-int}
\end{eqnarray}
where the differential $du$ integrates over $S^3$ and
\begin{equation}
\psi(u-u')=\mathcal{N}\, e^{-\frac{1}{2\kappa}\Delta L\left(\frac{u-u'}{\Delta
L}\right)^{2}} \;.
\label{hache}
\end{equation}  
$Q_M$ can be obtained by iterating a Chapman-Kolmogorov recurrence relation that relates polymers with $j$ and $j-1$ monomers, starting from an initial condition
\begin{equation}
Q_{0}(x,u,x_{0}, u_{0}) = \delta(x-x_{0})\delta(u-u_{0})\, I_{\mathscr{D}_{\rm Ad}} \;. 
\label{inic}
\end{equation}
As a result, when $j=M$, it is obtained
\begin{eqnarray} 
&& Q_{M}(x,u,x_0,u_0) = \int du'\,
 \psi(u-u') \,   e^{-\mu  \Delta L}  e^{-H(x,u)\Delta L} \, Q_{M-1}(x-u \Delta L,x_{0},u',u_{0})  \;.
\end{eqnarray} 
Expanding to first order in $\Delta L$ with finite $\kappa $, and taking the continuum limit, we arrive at the Fokker-Plank equation
\[
\partial_{L} Q = -\left[\mu  - \frac{\kappa}{\pi}\, \hat{L}^{2}_{u}+u_{\mu}\partial_{\mu}  
+ H(x,u) \right] Q   \;,
\]   
where $u_\mu \partial_\mu$ gets combined with $H(x,u)$ in Eq. \eqref{Hg}
to form the non-Abelian covariant derivative,
\begin{gather}
\left[\partial_L - (\kappa /\pi)\, \hat{L}^{2}_{u}+ \mu +  u_\mu \left( \partial_{\mu} - i\, {\rm Ad} \big(\Lambda_{\mu} \right) \big)\right] Q(x,u, x_0 , u_{0},L)=0 \;, 
\label{FPna}  \\ 
Q(x, u, x_0,u_0,0)= \delta(x-x_{0})\, \delta(u-u_{0})\, I_{\mathscr{D}_{\rm Ad}} 
\;.
\label{iniab1}
\end{gather}    
In the flexible limit (small stiffness), there is almost no correlation between the initial and final tangent vectors. The weak dependence on these directions allows us to  
consistently solve the equations by keeping the smaller angular momenta (see Refs.  \cite{GBO},  \cite{PhysRevLett.73.3235}).  In the present case, we get   
\begin{gather}
Q(x, u , x_0, u_0,L) \approx Q_0(x,x_0,L) \makebox[.5in]{,} \partial_L Q_0(x,x_0,L) = -O\, Q_0(x,x_0,L)   \;, 
\label{FPna-0} \\  
O = - c\, \left( \partial_{\mu} - i\, {\rm Ad} \big(\Lambda_{\mu} \big) \right)^2  + \mu  \makebox[.5in]{,}  Q_0(x,x_0,0)= \frac{1}{\Omega_3}\delta(x-x_{0}) \, I_d \;,
\label{iniab1-0}
\end{gather} 
where $c= \frac{\pi}{12 \kappa} \,$ and $\Omega_3$ is the solid angle on $S^3$.  Using this information in Eq. \eqref{pfun} yields,
\[  
\int d^4x\, du\; {\rm tr}\,  Q(x,x, u,u,L) \approx  {\rm Tr}\, \big(  e^{-LO} \big) \;.
\] 
In the second member, the trace is over the adjoint matrix indices and the spacetime coordinate $x$. Therefore, the loop sector is approximated by
\begin{eqnarray}  
Z_{\rm loops} [\Lambda]=   e^{  \int_{0}^{\infty}\frac{dL}{L}\;  \int  dv \,  {\rm tr}\, Q(v,v,L)  }
\approx e^{\, - {\rm Tr} \,  \ln O  } = ({\rm Det }\, O)^{-1}   \;,
\label{laste} 
\end{eqnarray} 
which can be represented by an effective complex field in the adjoint (see section \ref{eff-fields}). This is in contrast to the situation in Ref. \cite{GBO} where an ensemble formed by loops carrying internal degrees in a linear space parametrized by any set of complex numbers $z_C$, $C=1, \dots, N^2-1$ was analyzed. These variables label coherent states in an infinite dimensional space of color states \cite{ref26}. As a consequence, the effective representation of that ensemble required an infinite tower of fields carrying tensor products of adjoint irreps. It is interesting to note that the adjoint Wilson loop can also be related with internal degrees $z_C$ (cf. Eq. \eqref{z-var}) with the difference that they live in a nonlinear space given by the components of group-coherent states in the finite dimensional adjoint irrep. 

\section{Monopole fusion and effective Feynman diagrams}     
\label{frules}     
   
In Ref. \cite{GBO},  excluded volume effects and other interactions among monopoles were introduced as usual, namely by coupling them to external fields integrated with appropriate Gaussian  weights. The same steps could be done in Eq. \eqref{actionm}, but cubic terms would be missing in this formulation (see  the discussion in Ref. \cite{Proceed}). They will be relevant to drive ${\rm SU}(N) \to {\rm Z}(N)$  and  to describe the  observed first-order confining/deconfining phase transition when $N \geq 3$. In this section, they will be generated as a consequence of monopole fusion rules. 

Initially,  we shall replace $Z_{\rm loops}[\Lambda]$ in Eq. \eqref{funZ} by a general monopole sector $Z_{\rm m}[\Lambda] = Z_{\rm loops}[\Lambda] \,  Z_{\rm lines} [\Lambda]$. The first factor involves
 adjoint Wilson loops ${\mathcal W}_{\rm Ad}[\Lambda] $, giving rise to  a power of the functional determinant in Eq. \eqref{laste}, originated from loop copies needed to accommodate the matching rules (see section \ref{eff-fields}). The second factor is constructed in terms of holonomies $ {\rm Ad}(\Gamma[\Lambda])$ computed along open lines, forming (connected and disconnected) closed one-dimensional arrays. For a correct matching, they must be combined in a gauge-invariant way. 
 In this manner, when integrated over link variables, the lattice formulation of $Z_{\rm mix} [s_{\mu \nu}]$  
will receive contributions  from plaquettes distributed on: i) closed center vortex worldsurfaces generated by the dual YM term; ii) center vortices attached to loops; and iii) center vortices attached to one-dimensional arrays.  In the ensemble,  the lines $\gamma$ between given initial and final points $x_0$, $x$ will be weighted and integrated,  as we did with the Wilson loops in Eq. \eqref{ens-ad},
\[
 \int dL\, du \, du_0  \int [Dx]_{v_0 \, v}^{L}  \,  e^{- \int_0^{L}  ds\,  \left[  \frac{1}{2\kappa}\, \dot{u}_\mu
\dot{u}_\mu +
 \mu \right] } \,     {\rm Ad}(\Gamma[\Lambda])|_{A A'} \;. 
\] 
 The path-integral over shapes with fixed length $L$ gives the factor $Q(x,u,x_0,u_0,L)$ 
 treated in section \ref{first} (cf. Eq. \eqref{mat-con}).  In the flexible limit, using Eqs. \eqref{FPna}-\eqref{iniab1-0}, we obtain 
\begin{equation}
\int_0^\infty dL \, du\, du_0 \, Q(x , u ,x_0 , u_0,L) \sim  G( x,  x_0 )  \makebox[.5in]{,} O\, G(x,x_0) = \delta(x-x_0  ) \, I_{\mathscr{D}_{\rm Ad}} \;,
\end{equation}  
that is, a ($\Lambda$-dependent) Green's function $G(x , x_0)$  for every adjoint line.  As a result,  each array yields an  effective Feynman diagram.  By including coupling constants to measure the arrays' relative importance, the effective diagrams can be associated with a  
perturbative expansion of $Z_{\rm lines} [\Lambda] = 1+ C_{\rm lines}[\Lambda]$. In 4d, the relevant possibilities correspond to three and four fused lines. Therefore, we are interested in modeling  contributions
to $C_{\rm lines}[\Lambda]$ that involve blocks of the form 
\begin{eqnarray}
 C_n \propto  \int d^4x \, d^4x_0    \prod_{j=1 }^n  \int dL_{j} du^j  du_0^j  \int [Dx^{(j)}]_{v_0^j \, v^j}^{L_j}  \,  e^{- \int_0^{L_j}  ds_j\,  \left[  \frac{1}{2\kappa}\, \dot{u}^{(j)}_\mu
\dot{u}^{(j)}_\mu + 
 \mu \right] } \, D_n  \;, \label{eff-D} 
\label{Dgen-n-g} 
\end{eqnarray}
originated from all shapes and lengths of $n$ lines $\gamma_j$ ($n=3,4$) with common endpoints $x_0 $, $x$.

\subsection{Modeling $n$-line arrays}  

 For $n=3$, the gauge invariant $D_3$ could be given by
\[
D_3  = f_{ABC} \,  f_{A'B' C' }\;    {\rm Ad}(\Gamma_1[\Lambda])|_{A A'}\,  {\rm Ad}(\Gamma_2[\Lambda])|_{B B'} \, {\rm Ad}(\Gamma_3[\Lambda])|_{C C'} \;,
\]  
or with a combination of symmetric and antisymmetric structure constants in the place of $f_{ABC}$. 
In order to gain some insight about the possibilities, we note that the factors ${\mathcal W}^{(k)}_{\rm Ad}[\Lambda] $ in Eq. \eqref{fixmon} can be interpreted as monopole worldlines with non-Abelian d.o.f. Using the Petrov-Diakonov representation of the adjoint loop (Appendix \ref{gcoh-s}), we have
\begin{eqnarray} 
{\mathcal W}_{\rm Ad}[\Lambda] =  \int  [dg]_{\rm P} \, e^{  i \int_{\mathcal{C}}    dx_\mu \,    \left( g^{-1} \Lambda_\mu\, g 
+i\, g^{-1 }  \partial_\mu  g , \vec{\alpha}\cdot \vec{T} \right) } 
 \;,
\label{P-D-ad}
\end{eqnarray}
where $ \vec{\alpha}$ is a root. This leads to
\begin{eqnarray}
 Z_n [\Lambda] \approx \lefteqn{ \int [Dm]_n \prod_{k} \int[dg^{(k)}]_{\rm P} \,  e^{- \sum_k\int_0^{L_k}  ds_k\,  \left[  \frac{1}{2\kappa}\, \dot{u}^{(k)}_\mu 
\dot{u}^{(k)}_\mu +
 \mu \right] } } \nonumber \\
 && e^{   i \sum_{k} 
 \int_0^{L}  ds_k \,  u^{(k)}_\mu(s_k) \, \big(g^{-1} \Lambda_\mu(x(s))\, g 
+i\, g^{-1 }  \partial_\mu  g  , \, \vec{\alpha}\cdot \vec{T} \big)_k  }  \;.
\label{actionm}
\end{eqnarray}
The last factor can be thought of as a decoupled version of the group integrals originated from the monopole current $K_\mu = D_\nu(\tilde{L})\, {\cal F}_{\mu \nu}(Z) $ in configurations with chain defects, after using the identification  $ g^{(k)}(s_k) = \tilde{U}(x^{(k)}(s_k))$ (cf. Eqs. 
\eqref{newprop},   \eqref{Vrep} and
\eqref{VL}). This motivates the adoption of the gauge-invariant object
\begin{eqnarray}
D_n  =\int d\mu(g) d\mu(g_0) \, \langle g , \varepsilon_{1} |{\rm Ad}(\Gamma_1[\Lambda])  | g_0 , \varepsilon' _1 \rangle   \dots  \langle g , \varepsilon_{n} |{\rm Ad}(\Gamma_n[\Lambda])  | g_0 , \varepsilon' _{n} \rangle \;,
\label{Dgen-n} 
\end{eqnarray}    
where  $ |\varepsilon _j \rangle $, $ |\varepsilon'_j \rangle $ denote coherent reference states chosen as rotated root vectors (see Appendix \ref{gcoh-s}), as this choice allows us to make contact with the  monopole worldline interpretation.  In this regard, when $ |\varepsilon' _j \rangle =  |\varepsilon_j \rangle $,  we can write (cf. Eq. \eqref{global}) 
\begin{eqnarray}     
D_n  =\int d\mu(g) d\mu(g_0) \, \int \prod_j [dg^{(j)}(s_j)] \,  e^{ i \sum_j \int ds_j\,  ( g_j^\dagger \Lambda\, g_j + i g_j^\dagger  \dot{g}_j \, , 
\, X_j ) }  \makebox[.5in]{,} X_j = [E_j , E_j^\dagger]  \;.
\end{eqnarray} 
This is related to a monopole current
\begin{equation} 
K_\mu = 2\pi\, 2N\, \sum_j \tilde{U} X_j \, \tilde{U}^{-1}\, \int_{\mathcal{\gamma}_j}  dy_\mu\, \delta^{(4)}(x-y) \makebox[,5in]{,} \sum_j  X_j = 0 \;,
\label{X-con}
\end{equation}
with the identification $ g^{(j)}(s_j) = \tilde{U}(x^{(j)}(s_j))$, and $g_0$, $g$ given by 
the value of  $\tilde{U}$ at the line endpoints.
The last condition is a requirement for the covariant conservation of $K_\mu$ that we shall impose at each fusion point, thus generalizing the matching rule in the Cartan subalgebra $X_j = \vec{\delta}_j \cdot \vec{T}$, $\, \sum_j \vec{\delta}_j =0$,  discussed in Eq. \eqref{fusion-m}. 
In the flexible limit, path-integrating Eq. \eqref{Dgen-n-g}  over  $\gamma_j$, we obtain 
\begin{eqnarray}   
&&  C_n  \propto \int d^4x \, d^4x_0  \, \bar{F}^{\, \varepsilon_1 \dots \varepsilon_n  }_{A_1  \dots A_n}   F_{A'_1 \dots A'_n }^{\, \varepsilon'_1 \dots \varepsilon'_n  }\;    G(x , x_0)|_{A_1 A'_1}\, \dots\, G(x, x_0)|_{A_n A'_n} \;, \label{Cnline}    \\
&& F_{A_1  \dots A_n }^{\, \varepsilon_1 \dots  \varepsilon_n  } = \int d\mu(g)  \, | g, \varepsilon_{1} \rangle|_{A_1} \dots  | g, \varepsilon_{n} \rangle|_{A_n}   \;.
\label{Fal}  
\end{eqnarray} 
 
\subsection{Fusion of three monopoles}   
\label{three-fusion}

For three open worldlines, we have to compute
\begin{gather}
F_{A B C}^{\varepsilon_1 \varepsilon_2 \varepsilon_3  } =  \int d\mu(g)\,   | g , \varepsilon_{1} \rangle|_{A} \,  | g , \varepsilon_{2}  \rangle|_{B} \,  |g, \varepsilon_{3}\rangle|_{C} 
\makebox[.5in]{,} | g_0 , \varepsilon \rangle = R(g_0)\,   | \varepsilon \rangle \;, \\
 P_{A B C ; A' B' C'} = \int d\mu(g)  \,  R(g)|_{A A'} R(g)|_{B B'} R(g)|_{C C'} \makebox[.5in]{,} R(g)= {\rm Ad}(g)  \;.
\label{gp3}
\end{gather}   
The factor $R(g)|_{A A'} R(g)|_{B B'} $ acts on a tensor product space carrying a reducible representation. Its decomposition into irreps, complemented with the orthogonality relations 
\eqref{orth-rel}, leads to the desired integral.  For $N \geq 3$, there are seven irreps projected by $P_J$ \cite{Chivukula}-\cite{Cvitanovic},    
\begin{equation}
 \sum_{J} P_{[J]}^{AB, A'B'} =  \delta_{AA'} \delta_{BB'}  \makebox[.5in]{,}  
P_{[J]}^{AB,CD} P_{[K]}^{CD,A'B'} = \delta_{JK}\, P_{[J]}^{AB,A'B'} \;.
\end{equation}
They include a singlet $P_{[1]}$, and two projectors onto the adjoint $P_{[{\rm a}]}$, $P_{[{\rm s}]}$, with components
\begin{eqnarray}
 \frac{1}{N^2-1}\,  \delta_{AB} \delta_{A'B'}  \makebox[.5in]{,}  f_{ABC} f_{A'B'C}   \makebox[.5in]{,}
 \frac{N^2}{N^2-4}\, d_{ABC} d_{A'B'C}   \;,
\end{eqnarray} 
respectively. Note that in our conventions 
\begin{equation}
\{ T_A, T_B \} = \frac{1}{N^2}\,  \delta_{AB} I + d_{ABC} \, T_C  \makebox[.5in]{,} d_{ABC}\, d_{DBC} 
= 
\frac{N^2 - 4}{N^2}\, \delta_{AD} \;.
\end{equation} 
Hence,  we can write  
\begin{eqnarray}
 R(g)|_{A A'} R(g)|_{B B'} & = & R(g)|_{A \bar{A}} R(g)|_{B \bar{B}}\, \delta_{\bar{A}A'} \delta_{\bar{B}B'} \nonumber \\
& = & \frac{1}{N^2-1}\, \delta_{A B } \delta_{A'B'}  +  R(g)|_{A \bar{A}} R(g)|_{B \bar{B}}\, P_{\bar{A} \bar{B}; A'B'} + \dots\;, 
\label{prod-2} \\ 
P_{\bar{A} \bar{B}; A'B'}  = &&  f_{\bar{A} \bar{B} C} \, f_{A'B'C} + \frac{N^2}{N^2-4}\, d_{\bar{A} \bar{B} C} \, d_{A'B'C} \;, 
\end{eqnarray}
where the dots involve other irreps.  
Finally, the group invariance of the structure constants $f_{ABC}$ and $d_{ABC}$ yields
\begin{eqnarray}
&& P_{A B C ; A' B' C'}  = f_{A B C} f_{A'B'C'} + \frac{N^2}{N^2-4}\, d_{A B C} d_{A'B'C'}   \;, \nonumber \\
&& F_{A B C}^{\varepsilon_1 \varepsilon_2 \varepsilon_3  } =  f_{A B C}    (-i[E_{1}, E_{2}] , E_{3} )     
+  \frac{N^2}{N^2-4}\, d_{A B C} ( \{ E_{1}, E_{2}\} , E_{3} ) \;,
\label{Falpha}
\end{eqnarray}  
where $E = E|_A \, T_A$ is the Lie algebra element associated with $| \varepsilon \rangle$.  When this is replaced in Eq. \eqref{Cnline}, the cross terms with mixed symmetric and  antisymmetric  constants do not contribute, due to the symmetry of the product of Green's functions under the interchange $A \leftrightarrow B$, $A' \leftrightarrow B'$. Furthermore, if $ |\varepsilon' _j \rangle $ is an even ($+$) or an odd
($-$) permutation of $ |\varepsilon_j \rangle $, we get
\begin{eqnarray}
&& C^\pm_{\rm 3} \propto  \int d^4x \, d^4x_0  \, 
 G(x , x_0)|_{A A'}\, G(x , x_0)|_{B B'}\, G(x , x_0)|_{C C'}   \times  \nonumber \\
&&  [ \pm f_{A B C} f_{A' B' C'}\,  (-i[E_{1}, E_{2}] , E_{3} )^2
+ \frac{N^4}{(N^2-4)^2} \, d_{A B C} d_{A' B' C'} \, ( \{ E_{1}, E_{2}\}, E_{3} )^2 ]  \;.
\label{I3}
\end{eqnarray}
 
  The three-line Cartan matching only exists for $N \geq 3$, with
\begin{equation}
X_j = \vec{\delta}_j \cdot \vec{T}  \makebox[.3in]{,}  j=1,2,3  \makebox[.5in]{,} \vec{\delta}_1 + \vec{\delta}_2 +\vec{\delta}_3  =0 \;. 
\label{3-Cartan}
\end{equation}
 In this case, $E_j = E_{\delta_j}$ thus implying   
\[
 ([E_{\delta_1}, E_{\delta_2}] , E_{\delta_3}   )^2 = N^2_{ \delta_1 \delta_2}  (E_{\delta_1+ \delta_2}  , E_{\delta_3} )^2 = N^2_{ \delta_1 \delta_2}  \;,
\]  
and $ ( \{ E_{\delta_1}, E_{\delta_2}  \} , E_{\delta_3}  )^2 =  N^2_{ \delta_1 \delta_2} $. 
Now, the Weyl group for $\mathfrak{su}(N)$ acts as $S_N$, permuting the weights of the fundamental irrep \cite{Humphreys}. This  produces even but not odd permutations of three different roots. 
 This is the property underlying the two different contributions $C^\pm_{\rm 3}$. In Eq, \eqref{Dgen-n}, it is not possible to change variables in the group integral over $g_0$ to undone odd permutations.  In general, there are 
two independent combinations: the antisymmetric ($C^+_{\rm 3} -  C^-_{\rm 3}$)
\begin{eqnarray}
C^{\rm [a]}_{\rm 3-Cartan} \propto  \int d^4x \, d^4x_0 
\, G(x , x_0)|_{A A'}\, G(x , x_0)|_{B B'}\, G(x , x_0)|_{C C'} \,  N_{ \delta_1 \delta_2} ^2  f_{A B C}  f_{A' B' C'}   
\label{3Car}
\end{eqnarray}
 and the symmetric one, with $-i f_{ABC} \to d_{ABC} $.  
 
 Another natural matching type can be proposed for $N \geq 2 $ in the $\mathfrak{su}(2)$ subalgebras generated by $\frac{\vec{\alpha}\cdot \vec{T}}{\alpha^2}$,  $\frac{T_{\alpha}}{\sqrt{\alpha^2}}$,  
$\frac{T_{\bar{\alpha}}}{\sqrt{\alpha^2}}$.  
  As the directions $X_j$ have the same length, the solutions to
\begin{equation} 
X_\alpha^1 + X_\alpha^2 + X_\alpha^ 3 = 0 \;,
\label{3sum0}
\end{equation}  
 must be on the same plane and at angles of $2\pi/3$. Note that there is no common adjoint group action that can transform $X_\alpha^j$ into $ \vec{\delta}_j \cdot \vec{T}$, for $j=1,2,3$. Then, the former rule is physically inequivalent to the Cartan fusion type. The elements $X_\alpha^{j} = X_\alpha^{\theta_j}$,
\begin{equation}
X_\alpha^\theta =  \vec{\alpha}\cdot \vec{T} \cos \theta + \sqrt{\alpha^2}\,  
 T_{\alpha} \sin \theta = g(\theta) \,  \vec{\alpha}\cdot
\vec{T} \,g(\theta)^{-1}  \;,
\label{Xt}
\end{equation}
associated with $\theta_1 = 0$, $\theta_2 = \frac{2\pi}{3}$ and $\theta_3 = -\frac{2\pi}{3}$, satisfy Eq. \eqref{3sum0}.   
In this case, the rotated root  vectors are given by $E_j = E_\alpha^{\theta_j }$, 
\begin{eqnarray}
[E_\alpha^{ \theta} , E^\theta_{-\alpha} ] = X_\alpha^\theta \makebox[.5in]{,}  E_{\pm \alpha}^{ \theta} = g(\theta) \, E_{\pm \alpha } \, g(\theta)^{-1}    \;, \nonumber \\ 
( -i[  E_{\alpha}^{\theta_1},  E_{\alpha}^{\theta_2}] , E_{\alpha}^{\theta_3}  ) 
    =   \frac{3\sqrt{3}\, i}{4\sqrt{2}} \sqrt{\alpha^2} \makebox[.5in]{,} 
    ( \{  E_{\alpha}^{\theta_1},  E_{\alpha}^{\theta_2} \}  , E_{\alpha}^{\theta_3} ) 
     =   0  \;.
\end{eqnarray} 
This only leaves the antisymmetric part in Eqs. \eqref{Falpha} and \eqref{I3},  leading to $ F_{A B C}^{\, \theta_1 \theta_2 \theta_3}  = \frac{3\sqrt{3}}{4\sqrt{2}}\, i\sqrt{\alpha^2}  f_{A B C}$  and the corresponding contribution
\begin{eqnarray}
  C_{{\rm 3}-\mathfrak{su}(2)} \propto   \int d^4x \, d^4x_0  \,   G(x , x_0)|_{A A'}\, G(x , x_0)|_{B B'}\, G(x , x_0)|_{C C'} \, \alpha ^2   f_{A B C} \,  
f_ {A' B' C'}   \;.
\label{3su2}
\end{eqnarray}

\subsection{Fusion of four monopoles} 
 \label{four-fusion} 
 
For $n=4$,  we obtain
\begin{eqnarray}
F_{A \dots D}^{\, \varepsilon_1 \dots \varepsilon_4}   =     \frac{1}{N^2-1}\, \delta_{A B } \delta_{C D }  (E_1, E_2)
(E_3, E_4)  +   ( -i f_{A B E} ) (-i f_{C D E }) ( [E_1 , E_2 ] \, ,  [E_3 , E_4])  + \dots \;, \nonumber \\  \label{w-singlet}
\end{eqnarray}
where the dots involve $d_{A B E} 
\{ E_1 , E_2 \} $,  $ d_{C D E} \{ E_3 , E_4 \} $, and contributions due to other irreps. 
Using references $| \varepsilon_j \rangle = | \varepsilon_{\delta_j} \rangle $ associated with the matching rules in the Cartan sector (cf. Eq. \eqref{fusion-m}),  for the antisymmetric combination, we obtain the following  terms 
\begin{eqnarray}
C^{\rm [a]}_{\rm 4-Cartan} \propto  \int d^4x \, d^4x_0 
\, G(x , x_0)|_{A A'} G(x , x_0)|_{B B'} G(x , x_0)|_{C C'}  G(x , x_0)|_{D D'}  \nonumber \\
\times \, V_{\delta_1 \delta_2 , \delta_3 \delta_4 }^2 \,   f_{A B \bar{C}} \, f_{C D \bar{C} }  \, f_{A' B' \bar{D}} \, f_{C' D' \bar{D} } \;, 
\label{3Car4}
\end{eqnarray}
\[
V_{\delta_1 \delta_2 , \delta_3 \delta_4 }  = \left\{  \begin{array}{ll}
N_{\delta_1 \delta_2} N_{\delta_3 \delta_4}   \;,  &  ~~\vec{\delta}_1 + \vec{\delta}_2  \neq 0   \\
\vec{\delta}_1 \cdot \vec{\delta}_3  \;,  &  ~~\vec{\delta}_1 + \vec{\delta}_2  = 0   \;.  \end{array} \right. 
\]

\section{Effective models} 
\label{eff-fields} 
 
Now, we would like to combine the different results in an effective field description. Let us initially consider the ensemble of monopole loops in  Eq. \eqref{ens-ad}   approximated by Eq. \eqref{laste}.  Introducing  a Lie algebra-valued complex field $\zeta = \zeta|_A \, T_A $  with mass dimension one, i.e., a vector field $|\zeta \rangle$ with components $\zeta|_A$,  $A=1, \dots, N^2-1$, we have
\begin{eqnarray}  
({\rm Det }\, O)^{-1}  =  \int [\mathcal{D}\zeta][\mathcal{D}\bar{\zeta}] \, e^{-\int d^4x\, \langle \zeta | c^{-1} O | \zeta \rangle}  \;.
\end{eqnarray}  
Then, in this case, Eq. \eqref{funZ}   becomes

\begin{gather}  
  Z_{\rm mix} [s_{\mu \nu}] = \int [{\cal D}\Lambda_\mu]  [\mathcal{D}\zeta][\mathcal{D}\zeta^\dagger] \,   e^{-\int d^4x\,    \frac{1}{4g^2} \,   \left( F_{\mu \nu}(\Lambda ) - 2\pi s_{\mu \nu} \vec{\beta}_{\rm e} \cdot \vec{T}  \right)^2  }
  e^{- \int d^4x\,  
\left(    ( D_\mu \zeta^{\, \dagger} , 
 D_\mu \zeta ) + m^2   (\zeta^\dagger, \zeta )  \right)}  \;, 
\nonumber \\   
 m^2 = (12/\pi) \,  \mu \kappa  \makebox[.5in]{,}  D_\mu(\Lambda)\, \zeta  = \partial_{\mu} \zeta - i\, [\Lambda_{\mu} ,  \zeta ]\;.  
\label{Zmon} 
\end{gather}  
When monopole fusion is included, the partition function has the general form
\begin{equation}
  Z_{\rm mix} [s_{\mu \nu}] = \int [{\cal D}\Lambda_\mu] \,   e^{-\int d^4x\,    \frac{1}{4g^2} \,   \left( F_{\mu \nu}(\Lambda ) - 2\pi s_{\mu \nu} \vec{\beta}_{\rm e} \cdot \vec{T}  \right)^2  }
Z_{\rm m} [\Lambda]  \makebox[.5in]{,} Z_{\rm m}[\Lambda] = Z_{\rm loops}[\Lambda] \,  Z_{\rm lines} [\Lambda] \;.
\label{mixm}
\end{equation} 
 Relying on a single complex field $\zeta$, although we can write  
\begin{equation} 
G(x, x_0)_{A A'} \propto  \int [\mathcal{D}\zeta][\mathcal{D}\zeta^\dagger] \,  \zeta^\dagger(x)|_{A} \zeta(x_0)|_{A'}\, e^{- \int d^4x\,  
\left(  ( D_\mu \zeta^{\, \dagger} , 
 D_\mu \zeta ) + m^2  (\zeta^\dagger, \zeta )  \right)} \;,
 \end{equation}   
 the correlator in Eq. \eqref{3Car} cannot be reproduced.  
In fact, as there is no common group element that can orient $ \vec{\delta}_j \cdot \vec{T} $, $j=1,2,3$  along the same Cartan direction, each monopole line entering a fusion point must be associated with different internal degrees $\delta_j$. Accordingly, the loop types must also  be expanded, which in turn allows capturing the desired one-dimensional arrays by using  
 \begin{eqnarray}
 Z_{\rm m}[\Lambda] & = &  \int [\mathcal{D}\zeta][\mathcal{D}\bar{\zeta}] \, e^{-\int d^4x\, [  ( D_\mu \zeta_\alpha^{\, \dagger} , 
 D_\mu \zeta_\alpha ) + V_{\rm H}(\zeta) ]   } \;, 
 \label{zetalpha} \\  
    V_{\rm H}(\zeta) &= & m^2  (\zeta_\alpha^\dagger, \zeta_\alpha ) + \gamma_{\rm c}\,  N_{\delta_1 \delta_2} (\zeta_{\delta_3} , \zeta_{\delta_1} \wedge \zeta_{\delta_2} ) + {\rm c.c.} + \dots    \label{ZmC} \makebox[.3in]{,}  X \wedge Y \equiv -i\, [X ,  Y]  \;,
    \label{cubzet}
 \end{eqnarray} 
with the fields summed over positive roots $\vec{\alpha}$ and over roots $\vec{\delta}_j$ ($\vec{\delta}_1 + \vec{\delta}_2 +  \vec{\delta}_3 = 0$). For negative root indices $-\vec{\alpha}$, the notation $\zeta_{-\alpha} \equiv \zeta_\alpha^\dagger $ is understood. 
The dots 
involve the symmetric product $ \{ X , Y \} $ and constants $d_{ABC}$.  If only fusion types with
$ |\varepsilon' _j \rangle =  |\varepsilon_j \rangle $ were considered in  
Eq. \eqref{Dgen-n}, then the precise combination of vertices would be fixed by Eq. \eqref{Falpha}.

Expanding in $\gamma_{\rm c}$, we get a factor $Z_{\rm loops} [\Lambda] =\prod_\alpha Z_\alpha [\Lambda] = ({\rm Det }\, O)^{-N(N-1)/2 }$ times effective Feynman diagrams associated with three-line fusion. For instance, Eq. \eqref{3Car} is obtained from the average of the second order term
 \[
\int d^4x\,  d^4x_0\,  \gamma_{\rm c}^2 \, N^2_{\delta_1 \delta_2}\,  \big( \zeta_{\delta_3}^\dagger(x) ,  \zeta_{\delta_1}^\dagger(x) \wedge  \zeta_{\delta_2}^\dagger(x) \big) \, \big( \zeta_{\delta_3}(x_0) , \zeta_{\delta_1}(x_0) \wedge \zeta_{\delta_2}(x_0) \big)   \;. 
 \]     
 Now, let us include three-line fusion in the $\mathfrak{su}(2)$ subalgebras. To accomodate the matching condition \eqref{3sum0}, which involves generalized directions $X_\alpha^\theta$, one possibility is to further expand   
the loop types, labeling them with the different global orientations $X_{\xi} =  \xi\, \vec{\alpha}\cdot \vec{T} \, \xi^{-1}$. The loop contribution is then replaced by 
 \begin{equation}
Z_{\rm loops}[\Lambda]  =  e^{\sum_{\alpha} \ln Z_{\alpha} }  \to e^{\int d\mu(\xi) \ln Z_{\xi} } \;.
\end{equation}    
To avoid overcounting, the integral over the coset must be restricted. For every $\xi$, there is a $\xi'$ such that $X_{\xi'}= -X_{\xi} $. On the other hand, opposite points are already included in the loop orientations, so the integral is in fact over half the coset, 
\begin{equation}
Z_{\rm loops} [\Lambda] =  e^{ \mathscr{D}_{\rm Ad}/2 \ln Z_{\alpha} } =  ({\rm Det }\, O)^{-\frac{\mathscr{D}_{\rm Ad}}{2} } \;, 
\label{globalZ} 
\end{equation} 
thus leading to  $\mathscr{D}_{\rm Ad}$ real adjoint fields $\psi_A \in \mathfrak{su}(N)$.  
Like in the Cartan decomposition of a Lie basis (cf. \eqref{Tes}), the $N^2-1$ fields may be also organized as $\psi_\alpha$, $\psi_{\bar{\alpha}}$,  labeled by the positive roots $\vec{\alpha}$,  plus a sector $\psi_q$, $q=1, \dots, N-1$. In this manner, both fusion types are accommodated by the kinetic and potential terms
\begin{gather} 
  \frac{1}{2} (D_\mu \psi_A , 
 D_\mu \psi_A )   =  ( D_\mu \zeta_\alpha^{\, \dagger} , 
 D_\mu \zeta_\alpha ) + \frac{1}{2} (D_\mu \psi_q , 
 D_\mu \psi_q )   \nonumber \\
  V_{\rm H}(\psi) =  V_{\rm H}(\zeta)  +  \frac{m^2}{2}  (\psi_q , \psi_q ) + \gamma_{\mathfrak{su}(2)} \,  \vec{\alpha}|_q (\psi_q , \zeta_\alpha \wedge \zeta^\dagger_{\alpha} ) + \dots  \,,   \label{ZmC}  
\end{gather}
 where the complex fields are understood as $\zeta_{\pm\alpha} \equiv  (\psi_\alpha \pm i \psi_{\bar{\alpha}})/\sqrt{2}$.  When $\gamma_{\rm c }= \gamma_{\mathfrak{su}(2)}$, the ${\rm Ad}(SU(N))$-flavor symmetry of the loop sector is extended to the interactions, in which case,  
 \begin{eqnarray} 
 V_{\rm H}(\psi)  =   \frac{m^2}{2}  (\psi_A , \psi_A ) + \gamma \,  f_{ABC} \, (\psi_A , \psi_B \wedge \psi_C) + \dots   
 \label{3-cub}
 \end{eqnarray}  
  The remaining four-line fusion rules in Eq. \eqref{3Car4} are obtained from 
 \begin{equation}
N_{\delta_1 \delta_2} N_{\delta_3 \delta_4} (\zeta_{\delta_1} \wedge \zeta_{\delta_2} , \zeta_{\delta_3} \wedge \zeta_{\delta_4} )  + {\rm c.c.}  \makebox[.3in]{,} 
\vec{\alpha} \cdot \vec{\sigma} \, (\zeta_{\alpha}  \wedge \zeta_{\alpha}^\dagger , \zeta_{\sigma} \wedge \zeta^\dagger_{\sigma} )  \makebox[.3in]{,}    \vec{\alpha }|_q \, \vec{\alpha}|_p \, (\psi_q \wedge \zeta_{\alpha} ,  \zeta_{\alpha}^\dagger \wedge\psi_p )   \;. 
\label{q-terms}
\end{equation}
The first (second) term contributes to the case  $\vec{\delta}_1 + \vec{\delta}_2 \neq 0$ 
($\vec{\delta}_1 + \vec{\delta}_2 = 0$),  $\vec{\delta}_1  + \dots + \vec{\delta}_4 =0 $,  while the contribution of the third term is similar to that of the second with $\vec{\alpha} = \vec{\sigma} $.

As discussed throughout this work, the lattice version of 
 \begin{eqnarray}
  Z_{\rm mix} [s_{\mu \nu}] =  \int [{\cal D}\Lambda_\mu] [{\cal D }\psi ] \,   e^{-\int d^4x\,  \left[  \frac{1}{4g^2} \,   \left( F_{\mu \nu}(\Lambda ) - 2\pi s_{\mu \nu} \vec{\beta}_{\rm e} \cdot \vec{T}  \right)^2 +  \frac{1}{2} (D_\mu \psi_A , 
 D_\mu \psi_A ) +  V_{\rm H}(\psi)  \right]  }  \;,
 \label{Zmixp}
\end{eqnarray}
 normalized by $Z_{\rm mix} [0]$, is an average of center elements over percolating surfaces generated by the dual gauge sector $\Lambda_\mu$, that may be attached to loops and one-dimensional arrays generated by the $\psi$-sector. The various couplings  measure the abundance of each fusion type. A reduced model without the Cartan matching rules that involve different roots may have the form
\begin{equation}
 V_{\rm H}(\psi)  =  (\zeta_{\alpha}  \wedge \zeta_{\alpha}^\dagger - m\, \vec{\alpha} \cdot \vec{\psi}  )^2 + ( \vec{\alpha} \cdot \vec{\psi}   \wedge \zeta_{\alpha} - m\, \zeta_{\alpha} ,  \zeta^\dagger_{\alpha} \wedge  \vec{\alpha} \cdot \vec{\psi}   - m\, \zeta^\dagger_{\alpha}  ) \;.
 \label{point2}
\end{equation}
More generally, we could expand the squares and assign different couplings to the interaction terms.
The Higgs potential may also involve the symmetric product $ \{ X , Y \} $ and terms originated from other irreps, such as the singlet $(\zeta_\alpha^\dagger , \zeta_\alpha )(\zeta_\sigma^\dagger , \zeta_\sigma )$ (cf. Eqs. \eqref{Falpha}, \eqref{w-singlet}).  Among the alternatives, there is a natural ${\rm Ad}(SU(N))$ flavor-symmetric
one that encompasses all the couplings in Eqs. \eqref{cubzet}, \eqref{ZmC},  \eqref{q-terms},
 \begin{eqnarray}  
 V_{\rm H}(\psi)  =   \frac{m^2}{2}  (\psi_A , \psi_A ) + \frac{\gamma}{3} \,  f_{ABC} \, (\psi_A , \psi_B \wedge \psi_C) + \frac{\lambda}{4} \,  f_{ABC} f_{ADE} \, ( \psi_B \wedge \psi_C \, ,\, \psi_{D }\wedge \psi_{E} )  \;.  
 \label{piano}  
 \end{eqnarray}   
This model is analogous to that introduced in Ref. \cite{conf-qg}, with the difference that the quartic term  in that work was taken as $\frac{\lambda}{4} \,  ( \psi_A \wedge \psi_B)^2$. On fields of the form $\psi_A = \psi \, T_A $ both potentials coincide. Hence, we know that in the region $m^2 <  
(2/9)\,  \gamma^2/\lambda$ there is SSB, which corresponds to $ \mu <  (\pi/54) \, \gamma^2/\lambda \kappa$ (cf. Eq. \eqref{Zmon}). As our derivation is valid for positive stiffness $1/\kappa$, a negative $\mu$ certainly corresponds to a monopole condensate. This represents an ensemble where larger (precolating) monopole worldlines are favored. Nonetheless, because of the cubic terms, there is still the possibility of a monopole condensate with positive $m^2$. In this respect, 
for a given parameter choice, $ V_{\rm H}(\psi)$ can be written as a perfect square $V_{\rm H}(\psi)  =   \frac{1}{2} \, (m\, \psi_A - f_{ABC} \, \psi_B \wedge \psi_C)^2 $. In this case, as well as in  Eq. \eqref{point2}, the structure is similar to that present in $ N=1^\ast$ supersymmetric theories based on three complex adjoint Higgs fields \cite{Auzzi-Kumar}.

The obtained models have several common features that can be highlighted. The parameters can be chosen in order for the vacua manifolds to be given by
\begin{equation}
 \zeta_{\alpha}  \wedge \zeta_{\alpha}^\dagger = v\, \vec{\alpha} \cdot \vec{\psi} \makebox[.3in]{\rm }  \vec{\alpha} \cdot \vec{\psi}   \wedge \zeta_{\alpha} = v\, \zeta_{\alpha} \makebox[.7in]{\rm and}  f_{ABC} \, \psi_B \wedge \psi_C = v\, \psi_A \;,
\end{equation} 
respectively. The nontrivial solutions contain tuples $(\psi_1, \dots , \psi_{N^2-1})$, $\psi_A = v\, S T_A S^{-1}$, identified with points in  
${\rm Ad}({\rm SU}(N))$.   
When $ V_{\rm H}(\psi) $ is a perfect square ($v=m$), the nontrivial vacua are degenerate with the trivial point  $\psi_A =0$. However, for appropriate parameters, the degeneracy can be lifted. This triggers a phase where the dual gauge group ${\rm SU}(N)$ is broken to ${\rm Z}(N)$, which allows us to compute  $Z_{\rm mix} [s_{\mu \nu}]$  by means of a saddle point and collective modes. 
Therefore, in  the presence of the source $2\pi s_{\mu \nu} \vec{\beta}_{\rm e} \cdot \vec{T}$,
a flux tube with $N$-ality is induced. These models are also well-known to possess flux tube solutions with confined dual monopoles \cite{Hindmarsh-Kibble}-\cite{Marco-glue}. In particular, as the distance between a pair of adjoint quarks is increased, the saddle point will eventually favor  
string-breaking by screening the external sources with induced  dual monopoles, which get identified with valence gluons. Hence,  gluon confinement follows from the fact that the second homotopy group of ${\rm Ad}(SU(N))$ (a  compact group) is trivial. 
The difference-in-areas law for doubled pairs of ${\rm SU}(2)$ fundamental quarks can be similarly 
understood \cite{Proceed}. Furthermore, we could consider an observable formed by one adjoint and two fundamental holonomies with common endpoints, combined in a (chromoelectric) gauge-invariant way.  
This object could be used to 
calculate the hybrid potential for a quark-gluon-antiquark state in pure YM theory. 
The associated source in $ Z_{\rm mix} [s_{\mu \nu}]$  
contains a pair of surfaces carrying two different fundamental weights.  They are spanned between the adjoint and the fundamental lines. 
In accordance with the gluon interpretation, the induced saddle point will be a flux tube, running between the fundamental sources, with an induced dual monopole localized at the adjoint line.  

With respect to the L\"uscher corrections, the soft modes in a flavor nonsymmetric model will be given by the transverse fluctuations. This is welcomed, since the presence of additional gapless modes would modify
  the correction observed in lattice simulations up to $N=6$ \cite{Shif-qcd}, \cite{Teal}. YMH models that support flux tubes with $N$-ality and 
non-Abelian internal collective modes were constructed in Refs. \cite{Auzzi-Kumar} and \cite{GSY}.
They display $SO(3)_{\rm C-F}$ and $SU(N)_{\rm C-F}$ color-flavor locking, respectively. The   phenomenological effective models we derived may display a tensor product of $SO(3)_{\rm C-F}$ symmetries, one for each root, or ${\rm Ad}({\rm SU}(N))_{\rm C-F}$ symmetry.  Nevertheless,  in a YM context,
the  parameters would be related with a single scale, implying that possible non-Abelian degrees on the flux tube worldsheet are in fact frozen  \cite{Shif-qcd}.  For this reason, these phases would also be compatible with the observed universal corrections.

\section{Conclusions}  
\label{conc}

In this work, we initially considered a recently proposed gauge fixing in the continuum based on lattice center gauges, which induces a partition of the ${\rm SU}(N)$ YM path-integral into sectors with center vortex worldsurfaces and monopole worldlines. In this framework, we observed that physically inequivalent sectors are not only labeled by the location of defects but also by non-Abelian magnetic degrees of freedom.  The average of an observable involves two steps: a path-integral over general  fluctuations in each sector,  followed by an ensemble integration.

In the continuum, thin configurations amount to gauge fields $\mathscr{A}_\mu$ such that the field strength is localized on closed surfaces.   
 In this case, neither monopoles nor non-Abelian degrees affect the quark Wilson loop ${\mathcal W}_{\rm e}[\mathscr{A}]$. However, there are many possibilities for the ensemble measure, which dictates how to weight configurations when computing center-element averages. This measure should be obtained by taking the first step with the YM action, which is a difficult task. Instead, we suggested possible effects by considering a simple example based on a smoothed Gaussian version of the Wilson loop. In doing so, we observed that monopoles with non-Abelian d.o.f. get coupled to a dual non-Abelian gauge field  
$\Lambda_\mu$, in much the same way as in compact  ${\rm QED}(4)$. In addition,
the linking numbers of magnetic defects are encoded as a frustration in the action for $\Lambda_\mu$.
 
Motivated  by the above example, we proposed a measure to compute center-element averages in 4d ensembles of percolating center vortices and chains. 
In four dimensions, as center vortices are two-dimensional, the effective description would be related to a string field. However, in the condensed phase, it is known that a lattice string Higgs field model can be approximated by an Abelian gauge field representing the Goldstone modes \cite{Rey}. As a synthesis of the above physical inputs, and also guided by the 3d case, we associated a center 
vortex condensate in 4d with a Wilson action for a non-Abelian gauge field $\Lambda_\mu$ with frustration.   
This was implemented in a manner such that the lattice path-integral receives contributions from plaquettes distributed on closed surfaces. Moreover, they are accompanied by the center element that would be generated in the Wilson loop for quarks in representation ${\rm D} $.  For weaker dual coupling, larger and multiple surfaces are favored.  
In the next stage, monopoles were introduced by products of adjoint magnetic Wilson loop variables that single out plaquette  configurations distributed on surfaces attached in pairs to these loops. Using the Petrov-Diakonov representation, they were interpreted as monopole worldlines with non-Abelian d.o.f. Likewise, monopole fusion rules were introduced by means of gauge-invariant combinations of magnetic holonomies, involving three and four fused monopole lines. 

Finally, we 
integrated the monopole sector and showed that the large distance behavior  is given by a dual ${\rm SU}(N)$ YMH model with emergent adjoint Higgs fields.  
The field content depends on the physically inequivalent monopole loop types. Fusion rules 
in the Cartan and $\mathfrak{su}(2)$ subalgebras can be accomodated in models with $N^2-1$ real fields.  
When monopoles condense, the gauge group undergoes dual ${\rm SU}(N) \to {\rm Z}(N)$ SSB,  which makes it possible to capture the ensemble by means of a saddle point formed by flux tubes with $N$-ality and confined dual monopoles. If the parameters correspond to a flavor nonsymmetric model, the soft modes are only given by flux tube transverse fluctuations. In fact, this also occurs in the color-flavor locking phase, as the phenomenological parameters are expected to be originated from a single scale,  leaving no window for gapless non-Abelian modes \cite{Shif-qcd}. From this point of view, both possible scenarios are equally interesting, as they lead to 
the correct L\"uscher term observed up to $N=6$ in lattice simulations.  In order to narrow down the possibilities,  the various implied observables will be compared with lattice calculations in a future contribution. 

Thus, following the path proposed, we showed a possible mechanism to explain confining flux tubes and confined gluons as emergent objects in mixed ensembles of percolating center vortices and chains.

\section*{Acknowledgements}

The author thanks D. Tong for pointing out the similarity between our effective model and $N=1^\ast$ supersymmetric theories and G. M. Sim\~oes and D. R. Junior for discussions.
 The Conselho Nacional de Desenvolvimento Cient\'{\i}fico e Tecnol\'{o}gico (CNPq), the Coordena\c c\~ao de Aperfei\c coamento de Pessoal de N\'{\i}vel Superior (CAPES), and the Funda\c c\~{a}o de Amparo \`{a} Pesquisa do Estado do Rio de Janeiro (FAPERJ) are acknowledged for their financial support.

\appendix

 \section{Group-coherent states and holonomies}
\label{gcoh-s}

 \subsection{Group-coherent states} 

 Consider an irreducible $\mathscr{D}$-dimensional unitary representation over a vector space $\{ |\psi \rangle \}$.  The Lie algebra and group act according to
\begin{equation}
|\psi \rangle =  \left( \begin{array}{ccc}
\psi_1  \\
\vdots \\ 
\psi_\mathscr{D}  \end{array} \right) \makebox[.5in]{,}   |\psi \rangle 
 \to {\rm D}(Y) |\psi \rangle  \makebox[.5in]{,}  |\psi \rangle \to   {\rm D}(U) |\psi \rangle  \;. 
\label{rep-gen}
\end{equation}
Given a reference $|\phi \rangle$, $\langle \phi| \phi \rangle =1$, the invariance subgroup $H_\phi \subset G$ is defined by
\begin{equation} 
{\rm D}(h) |\phi \rangle = e^{i a(h )} |\phi \rangle \makebox[.5in]{,} h \in H_\phi \;. 
\label{inva} 
\end{equation}
 Coherent states of type $\{{\rm D}, |\phi \rangle\}$ are defined by  $|\xi, \phi \rangle = {\rm D}(\xi) |\phi \rangle $ \cite{Gilmore}, \cite{Perelemov}, after choosing a representative $\xi$ in the quotient $ G/H_\phi$, Then, as for every group element there is a unique decomposition $g = \xi h$, the action of $g$ becomes
\begin{equation}
{\rm D}(g)|\phi \rangle = e^{i a(h )} |\xi, \phi \rangle  \;.
\end{equation} 
Due to unitarity, the group invariance of the measure $d\mu(\xi)$ induced by the  
Haar measure $d\mu(g)$, and Schur's Lemma \cite{Humphreys}, the operator 
$O = \int d\mu(\xi)\,  |\xi, \phi \rangle \langle \xi, \phi | $ 
is proportional to the $\mathscr{D}\times \mathscr{D}$ identity matrix $I_{\mathscr D} $,  
\begin{equation} 
 \int d\mu(h) = 1   \makebox[.5in]{,} \int d\mu(\xi) = {\mathscr D}  \makebox[.5in]{,}  \langle \phi |\phi \rangle = 1 \makebox[.5in]{,} \int  d\mu(\xi) \, |\xi, \phi \rangle \langle \xi, \phi | = 
I_{\mathscr D}   \;. 
\label{over}
\end{equation}
That is, coherent states are overcomplete.

\subsection{Holonomies}  
\label{holo} 
 
The overcompleteness property  does not depend on the reference $|\phi \rangle$. However, 
in path-integral applications, some requirements must be considered.  
A reference state $|\phi \rangle$ such that the ``dynamical'' operator has a diagonal representation seems to be important to give meaning to the formal expressions \cite{Gilmore}, \cite{Klauder}.   Some irreps have weight vectors that enable a classical description, that is, a symplectic structure on the coset space. In particular, the highest weight vectors are among the favorable states \cite{Gilmore}-\cite{Simon}.  
The coherent state representation of the holonomy
\begin{equation}
 \Gamma[A]   =  P \left\{ e^{i \int_\gamma dx_\mu\,  A_{\mu}(x)  } \right\} 
\end{equation} 
is obtained by using the composition property with infinitesimal steps \cite{Kondo-coh,KOndo-98},   
\[  
{\rm D}(\Gamma[A]) =  (I_{\mathscr D}+ i\epsilon  {\rm D}(A(s_{M-1}) )\dots  (I_{\mathscr D}+ i\epsilon  {\rm D}(A(s_{0}) )  \makebox[.5in]{,} A(s) = \frac{dx_\mu}{ds} \, A_\mu(x(s)) \;,
\] 
 and then taking the continuum limit.  
     As usual, various completeness relations can be introduced. 
The reference $|\phi \rangle$ is chosen such that  the order $\epsilon$ contribution is nonzero \cite{Klauder}, with the second order providing a regularization \cite{Kondo-coh}. In this case, the factors can approximated by
\begin{eqnarray}
 1 + i\epsilon  \langle  \phi | {\rm D}(X_n) | \phi \rangle   
 \approx e^{ i\epsilon  \langle  \phi | {\rm D}(X_n) | \phi \rangle  } \makebox[.5in]{,} X_n= \xi_{n}^{\dagger}A(s_{n})\, \xi_{n} +i\xi_{n}^\dagger \,  \dot{\xi}_{n}  \;,
\label{approxi}
\end{eqnarray}
which leads to the representation   
\begin{eqnarray}  
\langle \xi, \phi |{\rm D}(\Gamma[A]) |\xi_0, \phi \rangle =  \int  [d\xi(s)]\, e^{ i \int ds\,  \langle \phi | {\rm D}(\xi^\dagger A\, \xi + i\xi^\dagger  \dot{\xi} ) |\phi\rangle }  \;,
\label{lineg}
\end{eqnarray} 
$  [d\xi]^{\,\xi_M}_{\xi_0}  = d\mu(\xi_{1}) \, d\mu(\xi_{2})  \dots  $, and the boundary conditions $\xi(0)= \xi_0 $, $ \xi(L)= \xi$.  Note also that
\begin{equation}
\langle \phi |{\rm D}( \xi^\dagger A \xi + i\xi^\dagger  \dot{\xi }) |\phi\rangle =   {\rm D}( A)|_{cd} \, z_d \bar{z}_c + \frac{i}{2} (\bar{z}_c \dot{z}_c - \dot{\bar{z}}_cz_c ) \;,
\label{z-var}
\end{equation}
where $a$ ranges from 1 to $ \mathscr{D}$ and $z_a(s)$ are the components of the coherent state $|z(s)\rangle =|\xi(s), \phi \rangle$.  
Following similar steps, using an identity based on the group,  we obtain
\begin{eqnarray}  
\langle g, \phi |{\rm D}(\Gamma[A]) |g_0, \phi \rangle =  \int  [dg(s)]\, e^{ i \int ds\,  \langle \phi | {\rm D}(g^\dagger A\, g + i g^\dagger  \dot{g} ) |\phi\rangle }   \;,
\label{linegg}
\end{eqnarray} 
with $ g(0)= g_0 $, $ g(L)= g$. The path $g(s)$ can be uniquely decomposed in the form $g(s) = \xi(s) h(s)$. Then, from Eq. \eqref{inva}, in the left-hand side of Eq. \eqref{linegg} we can replace $g \to \xi$, $g_0 \to \xi_0$, and include a factor $e^{i (a(0) - a(L))} $. 
Of course, this can be checked on the right-hand side by using
\begin{equation}
g^\dagger A\, g + i g^\dagger  \dot{g} = h^\dagger (\xi^\dagger A\, \xi + i \xi^\dagger  \dot{\xi} )h + + i h^\dagger  \dot{h}  \makebox[.5in]{,} \langle \phi| {\rm D} (h^\dagger(s) \dot{h}(s)) |\phi \rangle = i \dot{a}  \;.
\label{apunto}
\end{equation} 
In particular, as the Wilson loop is related to periodic boundary conditions, the coset and the group path-integrals have no relative factor,
\begin{eqnarray} 
&& {\mathcal W}_{\rm D}[A] = {\rm tr}\, {\rm D}(\Gamma[A]) =  \int [dg]_{\rm P}\,  e^{i \int ds\, \langle \phi |{\rm D}( g^\dagger A g + i g^\dagger  \dot{g }) |\phi\rangle }  \;.
\label{w-loop} 
\end{eqnarray}

\subsection{Maximal reference state}

 A general weight vector $|\phi_\lambda\rangle$ ($\langle \phi_\lambda| \phi_\lambda \rangle = 1$) satisfies
 \begin{equation}
{\rm D}(T_q) |\phi_\lambda \rangle = \vec{\lambda}|_q  |\phi_\lambda \rangle \;,
\label{weight-vector}
\end{equation}  
where $T_q$, $q=1, \dots, N-1$,  $[T_q, T_p]=0$, are independent elements generating the Cartan subalgebra. To compute
$\langle \phi_\lambda | {\rm D}(X) |\phi_\lambda\rangle $ for a general Lie algebra element $X \in \mathfrak{su}(N)$, we can expand it in the Cartan basis $T_q$, $E_{\alpha}$, $ E_{-\alpha}$, 
\begin{equation}
[T_q,E_\alpha]= \vec{\alpha}|_q\, E_\alpha \;.
\label{difw}  
\end{equation}
The step operators $E_{\alpha}$ are labelled by the positive roots $\vec{\alpha}$, which gives 
$N(N-1)/2$ possibilities\footnote{A weight is defined as positive if the last nonvanishing component is positive.}, while the hermitian generators can be identified with
\begin{eqnarray}
\{ T_A \} = \{ T_q , T_{\alpha}, T_{\bar{\alpha}} \} \makebox[.5in]{,} T_{\alpha}=\frac{1}{\sqrt{2}}(E_\alpha + E_{-\alpha})
\makebox[.5in]{,}
T_{\bar{\alpha}}=\frac{1}{\sqrt{2}i}(E_\alpha - E_{-\alpha})  \;.
\label{Tes} 
\end{eqnarray}
The remaining commutators are
\begin{equation}
[E_\alpha,E_{-\alpha}]= \vec{\alpha}|_q \, T_q \makebox[.3in]{,}  [E_\alpha, E_{\gamma}]=N_{\alpha \gamma}\, E_{\alpha+\gamma}
\makebox[.3in]{,}  \vec{\alpha}+\vec{\gamma} \neq 0 \;,
\end{equation}
where $N_{\alpha \gamma}=0$, if $\vec{\alpha}+\vec{\gamma}$ is not a root. 
If $\vec{\lambda}$ is the highest weight, then $|\phi_\lambda \rangle$ satisfies $E_\alpha |\phi_\lambda \rangle =0$, $ \langle \phi_\lambda | E_{-\alpha} =0$. In this case, in terms of the Killing form, we have
\begin{eqnarray}
&& \langle \phi_\lambda | {\rm D}(X) |\phi_\lambda\rangle = X^q \vec{\lambda}|_q 
=(X, \vec{\lambda}|_q T_q )  \;,
\label{cprope} \\  
&& \langle \phi_\lambda | {\rm D}(g^\dagger A\, g + i g^\dagger  \dot{g} ) |\phi_\lambda\rangle = ( g^\dagger A\, g + i g^\dagger  \dot{g } \, , \,  
\vec{\lambda}|_q T_q)    \;,
\label{PD-gen} 
\end{eqnarray}   
which leads to the Petrov-Diakonov representation of the Wilson loop  in Eq. \eqref{w-loop}  \cite{PD}.

\subsection{Adjoint representation}  
\label{adj-rep-s} 

For the adjoint representation, we have
\begin{equation}
 {\rm Ad}(Y)|_{AB} \zeta|_B \, T_A   =   [Y,\zeta]      \makebox[.5in]{,}    {\rm Ad}(U)|_{AB}\, \zeta|_B \, T_A = U \zeta U^{-1}  \;.
\end{equation}  
As the roots are formed by eigenvalues of the adjoint action of $T_q$ (cf. Eq. \eqref{difw}), they are weights of the  adjoint representation. In addition, the invariance subgroup,   
$ h E_\alpha h^{-1} = e^{i a(h )} E_\alpha $,
is the Cartan subgroup $h= e^{i\, \vec{c}\cdot \vec{T}}$, which gives $ a(h ) = 
 \vec{c}\cdot \vec{\alpha}$.  
Using the scalar product in  Eq. \eqref{Kf}, we get $\langle \zeta | Y \rangle = \bar{\zeta}|_A \, Y|_A = (\zeta^\dagger , Y )$.  Thus, for any reference $| \varepsilon \rangle = R(\xi) \, | \varepsilon_\alpha \rangle $ (i.e., $E = \xi E_\alpha \xi^{-1} $), the cyclicity of the Killing product yields
\[ 
\langle \varepsilon | {\rm Ad}(Y) | \varepsilon \rangle = ( E^\dagger , [Y , E ] ) = (Y , [ 
E , E^\dagger  ] ) = (Y , X  )  \makebox[.5in]{,} X = [E, E^\dagger ] = \xi  \, \vec{\alpha} \cdot \vec{T} \, \xi^{\dagger}\;.
\]
In terms of the rotated reference, Eq. \eqref{linegg} can be written in the form
\begin{eqnarray}  
 \langle g , \varepsilon|{\rm Ad}(\Gamma[\Lambda]) |g_0 , \varepsilon \rangle
& = & \int  [dg(s)] \,  e^{ i \int ds\,  ( g^\dagger \Lambda\, g + i g^\dagger  \dot{g} \, , 
\, X ) }    \;.  
\label{global}
\end{eqnarray}  
On the other hand, a coherent state reference $Z$ given by a combination of Cartan generators,
$[T_q, Z ] = 0 $,  cannot be used to derive a path-integral since
\[
\langle z | {\rm Ad}(Y) | z \rangle = ( Z^\dagger , [Y , Z ] ) = (Y , [ Z , Z^\dagger ] ) = 0 \;.
\]

\section{Orthogonality Relations}  
\label{ortog}
 
If ${\rm D}^{(i)}$ and ${\rm D}^{(j)}$ are unitary  irreps ($i\neq j$ label inequivalent irreps), then \cite{Hammermesh} 
\[
\int d\mu(g)\, {\rm D}^{(i)}(g)|_{a b}\,  {\rm D}^{(j)}(g^{-1})|_{  \beta \alpha}    =   \delta_{ij} \delta_{ a \alpha} \delta_{ b \beta}  \;.
\]
 In particular, for the adjoint, 
\begin{equation}
\int d\mu(g)\, R(g)|_{AB} \, R(g^{-1})|_{  B'A'}    =   \delta_{ A A'} \delta_{ B B'}  \;.
\label{orth-rel}
\end{equation}

\end{document}